\documentclass[a4paper,12pt]{article}

\usepackage{fancyhdr}
\usepackage[latin1]{inputenc}
\usepackage[usenames,dvipsnames]{pstricks}
\usepackage[permil]{overpic}
\usepackage{subfig}
\usepackage{amsmath}
\usepackage{amssymb}
\usepackage{enumerate}
\usepackage{placeins}
\usepackage{caption}
\usepackage{psfrag}
\usepackage{bm}
\usepackage{multirow}

\begin{document}

\pagestyle{fancy}

\begin{center}
{
\Large
\textbf{A nonlinear state-space approach \\ to hysteresis identification}

\vspace{1cm}
\normalsize
\begin{center}
\textbf{J.P. No\"el$^{1}$, A.F. Esfahani$^{1}$, G. Kerschen$^{2}$, J. Schoukens$^{1}$}
\end{center}

\begin{center}
$^{1}$ ELEC Department \\
Vrije Universiteit Brussel, Brussels, Belgium \\
\end{center}

\begin{center}
$^{2}$ Space Structures and Systems Laboratory \\
Aerospace and Mechanical Engineering Department \\
University of Li\`ege, Li\`ege, Belgium \\
\end{center}
\vspace{1cm}

\rule{0.85\linewidth}{.3pt}

\vspace{-0.5cm}

\begin{abstract}

Most studies tackling hysteresis identification in the technical literature follow white-box approaches, \textit{i.e.} they rely on the assumption that measured data obey a specific hysteretic model. Such an assumption may be a hard requirement to handle in real applications, since hysteresis is a highly individualistic nonlinear behaviour. The present paper adopts a black-box approach based on nonlinear state-space models to identify hysteresis dynamics. This approach is shown to provide a general framework to hysteresis identification, featuring flexibility and parsimony of representation. Nonlinear model terms are constructed as a multivariate polynomial in the state variables, and parameter estimation is performed by minimising weighted least-squares cost functions. Technical issues, including the selection of the model order and the polynomial degree, are discussed, and model validation is achieved in both broadband and sine conditions. The study is carried out numerically by exploiting synthetic data generated via the Bouc-Wen equations. 

\vspace{1cm}

\end{abstract}

\vspace{-0.5cm}

\rule{0.85\linewidth}{.3pt}

\vspace{1cm}
Corresponding author: Jean-Philippe Noël\\
Vrije Universiteit Brussel\\
ELEC Department, Pleinlaan 2\\
1050 Brussels, Belgium\\
\vspace{0.5cm}
Email: jeanoel@vub.ac.be

}
\end{center}

\newpage
\rhead{\thepage}
\section{Introduction}\label{Sec:Introduction}

Hysteresis is a phenomenology commonly encountered in a wide variety of engineering and science disciplines, ranging from solid mechanics, electromagnetism and aerodynamics~\cite{Hysteresis_Solids,Hysteresis_EM_Book,Hysteresis_Aero} to biology, ecology and psychology~\cite{Hysteresis_Biology,Hysteresis_Ecology,Hysteresis_Psychology}. In structural dynamics, hysteresis is mostly featured in joints, where it results from friction between assembled parts~\cite{Friction_Joints}. The defining property of a hysteretic system is the persistence of an input-output loop as the input frequency approaches zero~\cite{Bernstein_Hysteresis1}. Hysteretic systems are inherently nonlinear, as the quasi-static existence of a loop requires an input-output phase shift different from 0 and 180 degrees, which are the only two options offered by linear theory. The root cause of hysteresis is multistability~\cite{Bernstein_Hysteresis2}. A hysteretic system possesses multiple stable equilibria, attracting the output depending on the input history. In this sense, it is appropriate to refer hysteresis as system nonlinear memory. 

The identification of hysteresis is challenging, primarily because it is a dynamic kind of nonlinearity governed by internal variables, which are not measurable. Most studies tackling hysteresis identification in the technical literature follow white-box approaches, \textit{i.e.} they rely on the assumption that measured data obey a specific hysteretic model~\cite{Hassani_Hysteresis}. The Bouc-Wen model was identified in numerous works, in particular using optimisation techniques such as evolutionary~\cite{Charalampakis_Hysteresis_Evolution,Worden_Hysteresis_DiffEvolution} and particle swarm~\cite{Charalampakis_Hysteresis_Swarm} algorithms. In Refs.~\cite{Worden_Hysteresis_Bayes1,Worden_Hysteresis_Bayes2,Ortiz_Hysteresis_Bayes}, a Bayesian framework was exploited to quantify uncertainty in Bouc-Wen identification. Specialised NARX~\cite{Worden_HysteresisNARX_IMAC2012}, neural network~\cite{Xie_BoucWen_NeuralNets} and Hammerstein~\cite{Giri_Hysteresis} models were also developed to address Bouc-Wen systems. The experimental identification of other hysteresis models, like the Preisach equations and a stochastic Iwan description of friction, is reported in Refs.~\cite{Ktena_Preisach,Mignolet_Iwan}, respectively. Assuming a specific model structure in a white-box philosophy may however be a hard requirement to handle in real applications, since hysteresis is a highly individualistic nonlinear behaviour. 

In this paper, a black-box approach based on nonlinear state-space models is adopted to identify hysteresis dynamics. State-space identification is a powerful way to experimentally model nonlinear systems. A literature survey shows that systems as diverse as a magneto-rheological damper~\cite{Paduart_PNLSS}, a wet-clutch device~\cite{Widanage_PNLSS_Clutch}, a glucoregulatory system~\cite{Marconato_PNLSS_Gluco}, or a Li-Ion battery~\cite{Relan_PNLSS_Battery} were successfully identified using nonlinear state-space models. The approach proposed in the present paper exploits the great flexibility of a state-space representation to establish a general framework to hysteresis identification, which makes no use of a priori assumptions. Physical insight into the system behaviour can also be retrieved, ensuring a reasonable parsimony of the derived model. Nonlinear model terms are constructed as a multivariate polynomial in the state variables, and parameter estimation is performed by minimising weighted least-squares cost functions in the frequency domain. Technical issues, including the selection of the model order and the polynomial degree, are discussed, and model validation is achieved in both broadband and sine conditions. The present study is carried out numerically by exploiting synthetic data generated via the Bouc-Wen equations. However, it is emphasised that the Bouc-Wen nature of the data will not be exploited in the identification process. 

\newpage
The paper is organised as follows. The synthesis of input-output data is described in Section~\ref{Sec:Simulation}, where noise assumptions are also stated. A nonparametric study of the nonlinear distortions affecting the generated data is conducted in Section~\ref{Sec:NonParam}, and parametric modelling in state space is carried out in Section~\ref{Sec:PNLSS}. Model validation is eventually achieved in Sections~\ref{Sec:PNLSS} and~\ref{Sec:Validation}, and concluding remarks are formulated in Section~\ref{Sec:Conclusion}.

\section{Synthetic generation of hysteretic data}\label{Sec:Simulation}

The synthesis of noisy data exhibiting hysteresis behaviour is carried out in this section by combining the Bouc-Wen differential equations (Section~\ref{Sec:BoucWen}), multisine excitation signals (Section~\ref{Sec:Excitation}) and the Newmark integration rules (Section~\ref{Sec:Integration}). Noise assumptions are finally discussed in Section~\ref{Sec:Noise}.

\subsection{The Bouc-Wen model of hysteresis}\label{Sec:BoucWen}

The Bouc-Wen model~\cite{Bouc_Hysteresis_ENOC1967,Wen_Hysteresis} has been intensively exploited during the last decades to represent hysteretic effects in structural dynamics, especially in the case of random vibrations. Extensive literature reviews about phenomenological and applied aspects related to Bouc-Wen modelling can be found in Refs.~\cite{Ismail_Hysteresis_Survey,Ikhouane_BookHysteresis}. In this work, a Bouc-Wen system with a single degree of freedom is considered to demonstrate the applicability of state-space models to hysteresis identification. In that respect, multi-degree-of-freedom systems are out of the scope of the paper.  

The vibrations of a single-degree-of-freedom Bouc-Wen system are governed by Newton's law of dynamics written in the form~\cite{Wen_Hysteresis}
\begin{equation}
m_{L}\:\ddot{y}(t) + r(y,\dot{y}) + z(y,\dot{y}) = u(t) ,
\label{Eq:EOMBW}
\end{equation}
where $m_{L}$ is the mass constant, $y$ the displacement, $u$ the external force, and where an over-dot indicates a derivative with respect to the time variable $t$. The total restoring force in the system is composed of a static nonlinear term $r(y,\dot{y})$, which only depends on the instantaneous values of the displacement $y(t)$ and velocity $\dot{y}(t)$, and of a dynamic, \textit{i.e.} history-dependent, nonlinear term $z(y,\dot{y})$, which encodes the hysteretic memory of the system. In the present study, the static restoring force contribution is assumed to be linear, that is
\begin{equation}
r(y,\dot{y}) = k_{L}\:y + c_{L}\:\dot{y},
\label{Eq:EOMBW_Static}
\end{equation}
where $k_{L}$ and $c_{L}$ are the linear stiffness and viscous damping coefficients, respectively. The hysteretic force $z(y,\dot{y})$ obeys the first-order differential equation 
\begin{equation}
\dot{z}(y,\dot{y}) = \alpha\:\dot{y} - \beta \left(\gamma \: \left|\dot{y}\right| \left|z\right|^{\nu-1} z + \delta \: \dot{y} \left|z\right|^{\nu} \right) ,
\label{Eq:EOMBW_Hysteretic}
\end{equation}
where the five Bouc-Wen parameters $\alpha$, $\beta$, $\gamma$, $\delta$ and $\nu$ are used to tune the shape and the smoothness of the system hysteresis loop. Note that the variable $z$ is not measurable, which may complicate the formulation of an identification problem. Another difficulty is that Eq.~(\ref{Eq:EOMBW_Hysteretic}) is a nonlinear relation in the parameter $\nu$. These two issues will be addressed in Section~\ref{Sec:PNLSS_Results} through a black-box state-space modelling approach. 

\begin{figure}[p]
\begin{center}
\begin{tabular}{c c}
\subfloat[]{\label{Fig:IOLoop}\includegraphics[width=75mm]{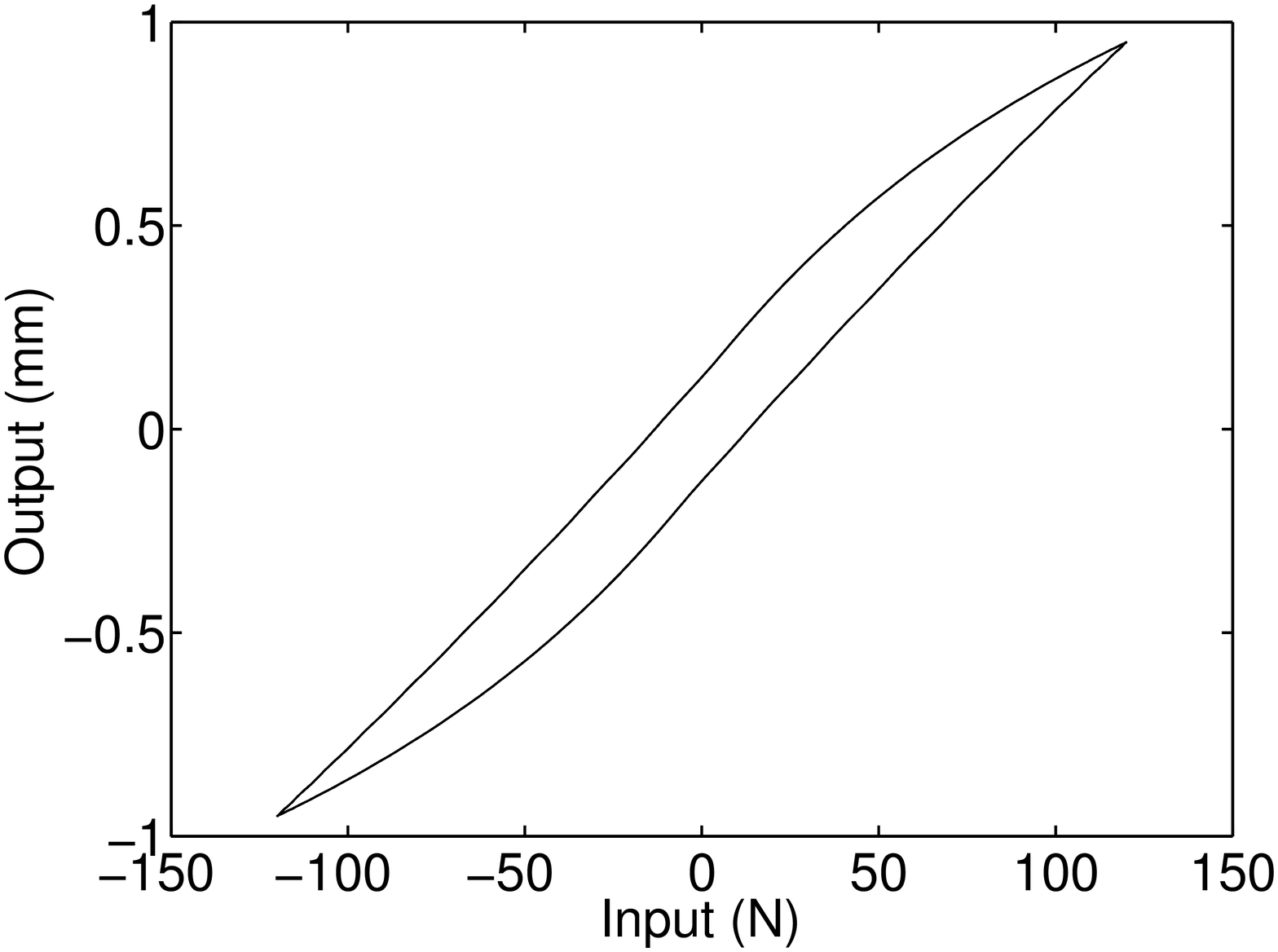}} &
\subfloat[]{\label{Fig:IONoLoop}\includegraphics[width=75mm]{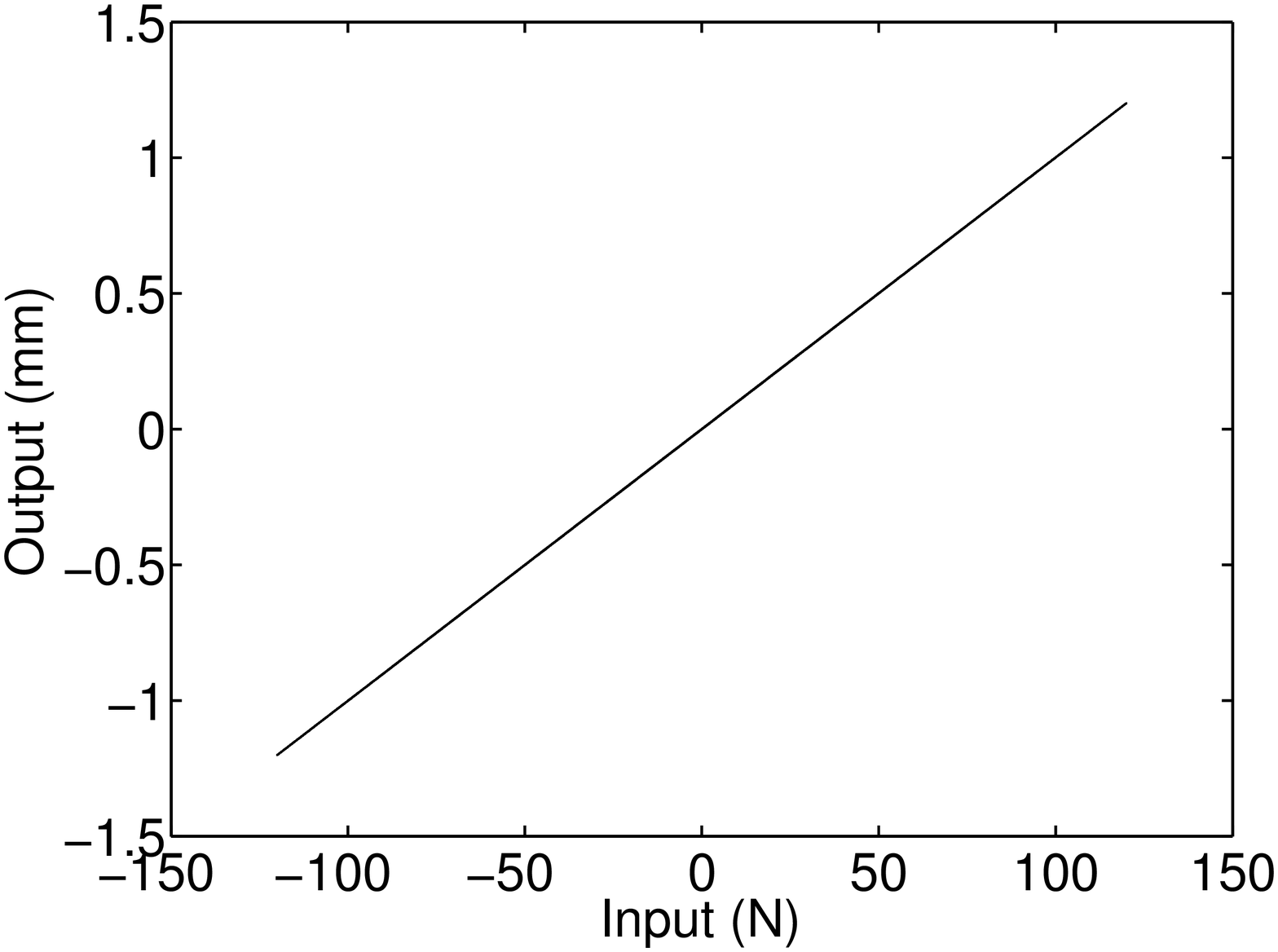}} \\
\end{tabular}
\caption{Hysteresis loop in the system input-output plane for quasi-static forcing conditions. (a) Non-degenerate loop obtained for the parameters in Table~\ref{Table:BWParams}; (b) linear behaviour retrieved when setting the $\beta$ parameter to 0.}
\label{Fig:IOHysteresisLoop}
\end{center}
\end{figure}

Table~\ref{Table:BWParams} lists the values of the physical parameters selected in this work. The linear modal parameters deduced from $m_{L}$, $c_{L}$ and $k_{L}$ are given in Table~\ref{Table:ModalParams}. Fig.~\ref{Fig:IOHysteresisLoop}~(a) illustrates the existence of a non-degenerate loop in the system input-output plane for quasi-static forcing conditions. In comparison, by setting the $\beta$ parameter to 0, a linear behaviour is retrieved in Fig.~\ref{Fig:IOHysteresisLoop}~(b). The excitation $u(t)$ in these two figures is a sine wave with a frequency of 1 $Hz$ and an amplitude of 120 $N$. The response exhibits no initial condition transients as it is depicted over 10 cycles in steady state. 

\begin{table}[p]
\centering
\begin{tabular}{c c c c c c c c c c}
\hline
Parameter & $m_{L}$ & $c_{L}$ & $k_{L}$ & $\alpha$ & $\beta$ & $\gamma$ & $\delta$ & $\nu$ \\
Value (in SI unit) & 2 & 10 & $5\:10^{4}$ & $5\:10^{4}$ & $1\:10^{3}$ & 0.8 & -1.1 & 1 \\
\hline
\end{tabular}
\caption{Physical parameters of the Bouc-Wen system.} 
\label{Table:BWParams}
\end{table}

\begin{table}[p]
\centering
\begin{tabular}{c c c}
\hline
Parameter & Natural frequency $\omega_{0}$ ($Hz$) & Damping ratio $\zeta$ ($\%$) \\
Value & 35.59 & 1.12 \\
\hline
\end{tabular}
\caption{Linear modal parameters of the Bouc-Wen system.} 
\label{Table:ModalParams}
\end{table}

\subsection{Excitation signal}\label{Sec:Excitation}

Multisine excitations will be considered throughout the identification process. A multisine is a periodic signal with a user-defined amplitude spectrum and randomly-chosen phases. It is also known as a pseudo-random signal, since it features a random appearance in the time domain and deterministic amplitudes in the frequency domain. Multisines are attractive because they are periodic, broadband, and low-crest-factor signals~\cite{JSchoukens_Book}. In particular, periodicity allows leakage to be eliminated in frequency-domain identification and the covariance matrix of the noise disturbances to be estimated directly from data.

Formally, a multisine input $u(t)$ is defined by means of a sum of sine waves with related frequencies as~\cite{JSchoukens_Book}
\begin{equation}
u(t) = N^{-1/2} \displaystyle \sum_{k=-N/2+1}^{N/2-1} U_{k} \: e^{j\left( 2\pi \:k \: \frac{f_{s}}{N} \: t + \phi_{k} \right)} ,
\label{Eq:Multisine}
\end{equation}
where $U_{k} = U_{-k}$, $U_{0} = 0$, and $\phi_{k} = -\phi_{-k}$. $N$ is the number of time samples, $N^{-1/2}$ a scaling factor, $f_{s}$ the sampling frequency, and $j$ the imaginary unit. The amplitudes $U_{k}$ are controlled by the user to meet a desired spectrum (see Sections~\ref{Sec:NonParam} and~\ref{Sec:PNLSS_Results}), and the phases $\phi_{k}$ are drawn from an uniform distribution on $\left[0,\:2\pi\right)$. The signal $u(t)$ in Eq.~(\ref{Eq:Multisine}) is asymptotically normally distributed in the time domain as $N$ tends to infinity. 

\subsection{Time integration}\label{Sec:Integration}

The Bouc-Wen dynamics in Eqs.~(\ref{Eq:EOMBW}) and~(\ref{Eq:EOMBW_Hysteretic}) can be effectively integrated in time using a Newmark method. Newmark integration relies on one-step-ahead approximations of the velocity and displacement fields obtained by applying Taylor expansion and numerical quadrature techniques~\cite{Geradin_Book}. Denoting by $h$ the integration time step, these approximation relations write
\begin{equation}
\begin{array}{c}
\dot{y}(t+h) = \dot{y}(t) + \left(1-a\right) \: h \: \ddot{y}(t) + a \: h \: \ddot{y}(t+h) \\
y(t+h) = y(t) + h \: \dot{y}(t) + \left(\frac{1}{2} - b \right) \: h^{2} \: \ddot{y}(t) + b \: h^{2} \: \ddot{y}(t+h) .\\
\end{array}
\label{Eq:Newmark}
\end{equation}
Parameters $a$ and $b$ are typically set to 0.5 and 0.25, respectively. Eqs.~(\ref{Eq:Newmark}) are herein enriched with an integration formula for the variable $z(t)$, which takes the form
\begin{equation}
z(t+h) = z(t) + \left(1-c\right) \: h \: \dot{z}(t) + c \: h \: \dot{z}(t+h) ,
\label{Eq:NewmarkBW}
\end{equation}
where $c$, similarly to $a$, is set to 0.5, since the two parameters play analogous roles in Eq.~(\ref{Eq:Newmark}) and Eq.~(\ref{Eq:NewmarkBW}), respectively. The sensitivity of the Newmark integration to parameter $c$ was evaluated, without noticing any appreciable alteration of the calculated time histories for values different than 0.5. Based on Eqs.~(\ref{Eq:Newmark}) and~(\ref{Eq:NewmarkBW}), a Newmark scheme proceeds in two steps. First, predictions of $\dot{y}(t+h)$, $y(t+h)$ and $z(t+h)$ are calculated assuming $\ddot{y}(t+h) = 0$ and $\dot{z}(t+h) = 0$. Second, the initial predictors are corrected via Newton-Raphson iterations so as to satisfy the dynamic equilibria in Eqs.~(\ref{Eq:EOMBW}) and~(\ref{Eq:EOMBW_Hysteretic}).

\begin{figure}[ht]
\begin{center}
\begin{tabular}{c c}
\subfloat[]{\label{Fig:TimeOutput}\includegraphics[width=75mm]{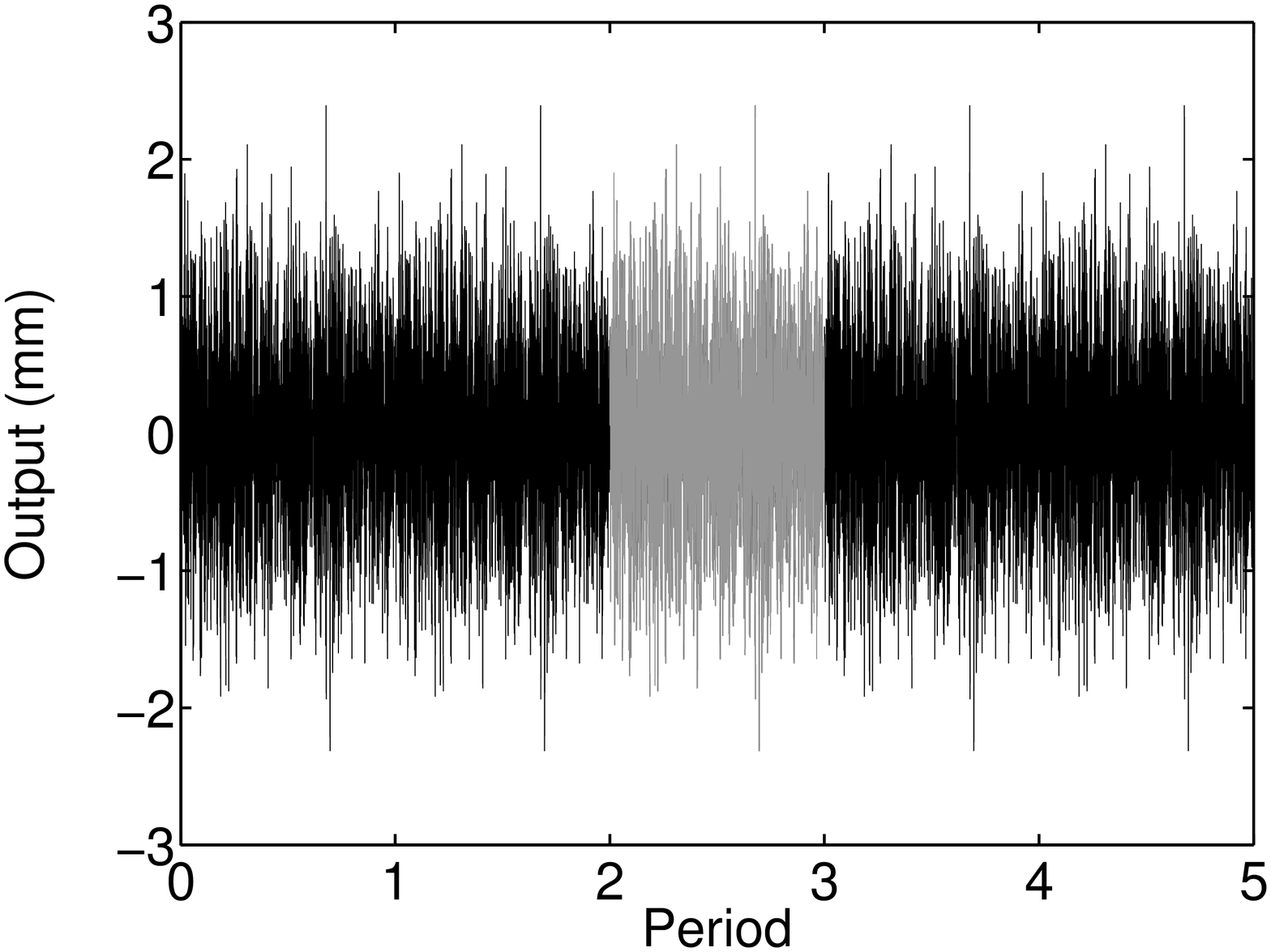}} &
\subfloat[]{\label{Fig:TransientDecay}\includegraphics[width=75mm]{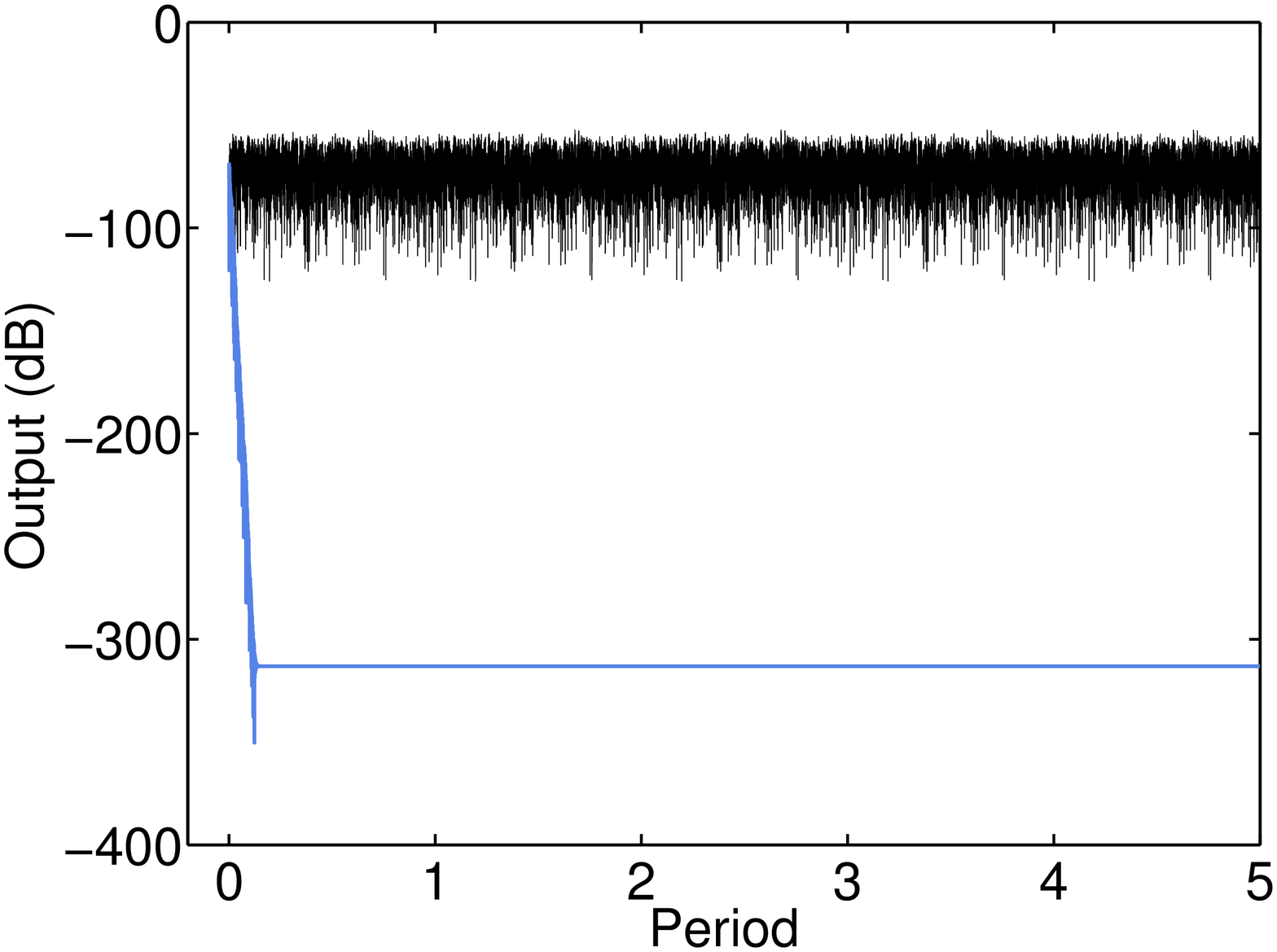}} \\
\end{tabular}
\caption{System output calculated in response to a multisine input band-limited in 5 -- 150 $Hz$. (a) Output over 5 periods, with one specific period highlighted in grey; (b) output in logarithmic scaling (in black) and decay of the transient response (in blue).}
\label{Fig:TimeIntegration}
\end{center}
\end{figure}

The sampling rate during integration, \textit{i.e.} $1/h$, is set to 15000 $Hz$. For identification use, synthesised time histories are low-pass filtered and downsampled to 750 $Hz$. Fig.~\ref{Fig:TimeIntegration}~(a) displays the system output calculated in response to a multisine input for which all frequencies in the 5 -- 150 $Hz$ band are excited. The root-mean-squared (RMS) amplitude of the input is 50 $N$ and 5 output periods are simulated. The exponential decay of the system transient response is plotted in Fig.~\ref{Fig:TimeIntegration}~(b) by subtracting the last synthesised period from the entire time record. This graph indicates that transients due to initial conditions only affect the first period of measurement, and that the applied periodic input results in a periodic output. It also demonstrates the high accuracy of the Newmark integration, as the transient response reaches the Matlab precision of -313 $dB$ in steady state. 

\subsection{Noise assumptions}\label{Sec:Noise}

White Gaussian noise is added to the synthesised output measurements $y(t)$ considering a signal-to-noise ratio (SNR) of 40 $dB$. The input time series $u(t)$ is assumed to be noiseless, \textit{i.e.} observed without errors and independent of the output noise. In practice, electromagnetic shakers typically yield a SNR of 60 to 80 $dB$, which is coherent with a noise-free assumption. If the input noise disturbances are otherwise too important, measurements can be averaged over multiple periods.

\section{Nonparametric analysis of nonlinear distortions}\label{Sec:NonParam}

\begin{figure}[ht]
\begin{center}
\includegraphics[width=150mm]{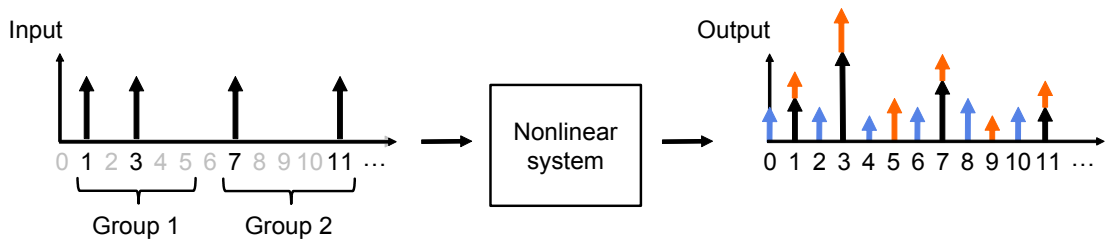}
\caption{Multisine input spectrum with well-selected measurement (in black) and detection (in grey) lines, and corresponding output spectrum where odd (in orange) and even (in blue) nonlinear distortions are quantified and separated.}
\label{Fig:NonParam}
\end{center}
\end{figure}

In this section, a simple testing procedure is employed to gain rapid insight into the nonlinear distortions observed in output data. The procedure requires no user interaction, and no parametric modelling effort. It allows an objective, \textit{i.e.} a quantitative, detection of nonlinear behaviour and separates distortions originating from odd and even nonlinearities. The basic idea of the approach is to design a multisine excitation signal comprising odd frequencies only, called measurement lines, and to assess nonlinear distortions by measuring the output level at the nonexcited frequencies, called detection lines~\cite{Schoukens_LinNL}. More specifically, Fig.~\ref{Fig:NonParam} shows that the adopted input amplitude spectrum possesses no even frequencies, serving as even detection lines. In addition, odd excited frequencies are grouped into sets of successive lines (for instance, 1 -- 3 -- 5 and 7 -- 9 -- 11), and one frequency is randomly rejected from each group to function as an odd detection line (for instance, 5 and 9)~\cite{JSchoukens_BookExercises}. This specific choice of input spectrum permits the following classification of the output spectrum contributions in Fig.~\ref{Fig:NonParam}, assuming steady-state conditions~\cite{Schoukens_BLA}:
\begin{itemize}
	\item at the measurement lines, linear dynamics (in black) and odd nonlinear distortions (in orange) appear ;
	\item at the odd detection lines, only odd nonlinear distortions (in orange) are visible ;
	\item at the even detection lines, only even nonlinear distortions (in blue) emerge.
\end{itemize}

It should be noted that, in practice, interactions between the actuator and the system can disturb the desired input spectrum, and hence jeopardise this nonparametric distortion analysis. This situation is addressed in Ref.~\cite{Pintelon_ClosedLoop} by taking advantage of the knowledge of the reference input signal, and in Ref.~\cite{Schoukens_NonIdeal} where first-order corrections to the distortion analysis are calculated. Moreover, Fig.~\ref{Fig:NonParam} shows that, at the excited frequencies (1 -- 3 -- 7 -- 11), it is not possible to separate the linear contributions from the odd nonlinear distortions. This makes odd nonlinear distortions potentially more harmful than even distortions. However, the level of observed distortions at the unexcited frequencies can be extrapolated to the excited frequencies if a randomised frequency grid is adopted, as explained in Refs.~\cite{Pintelon_RobustBLA,Schoukens_RobustBLA}.

\begin{figure}[ht]
\begin{center}
\begin{tabular}{c c}
\subfloat[]{\label{Fig:NonParam1}\includegraphics[width=75mm]{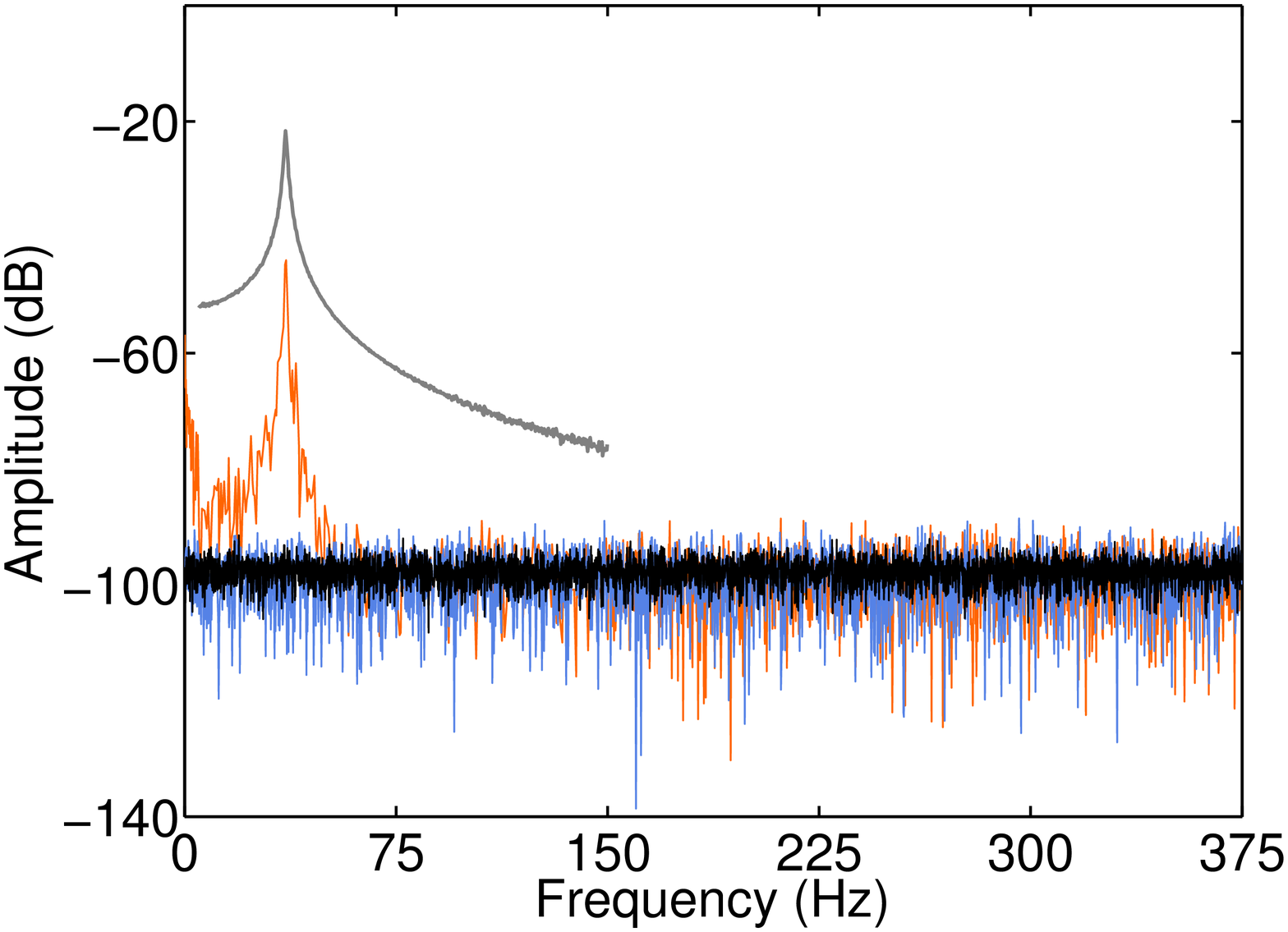}} &
\subfloat[]{\label{Fig:NonParam10}\includegraphics[width=75mm]{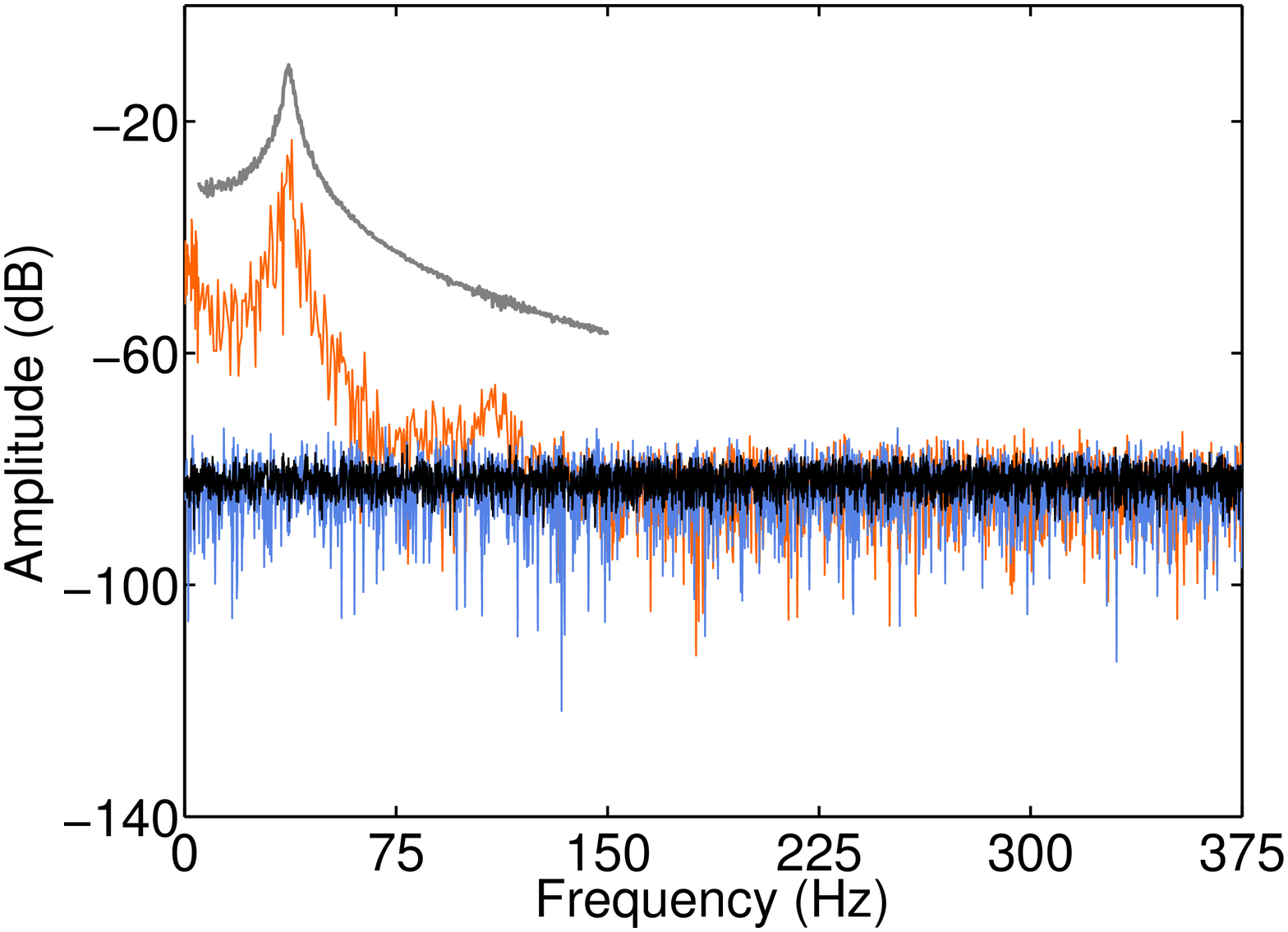}} \\
\subfloat[]{\label{Fig:NonParam25}\includegraphics[width=75mm]{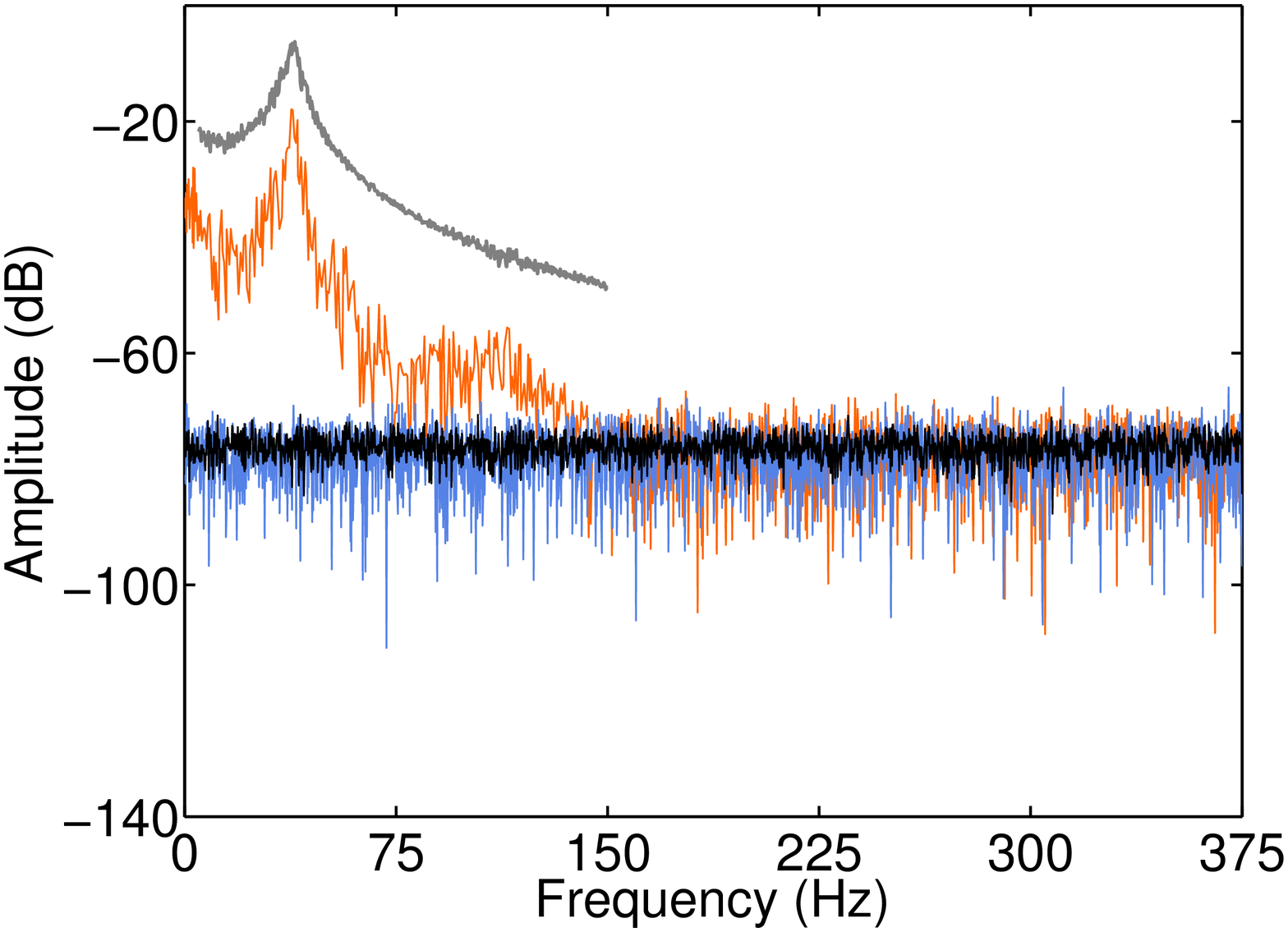}} &
\subfloat[]{\label{Fig:NonParam50}\includegraphics[width=75mm]{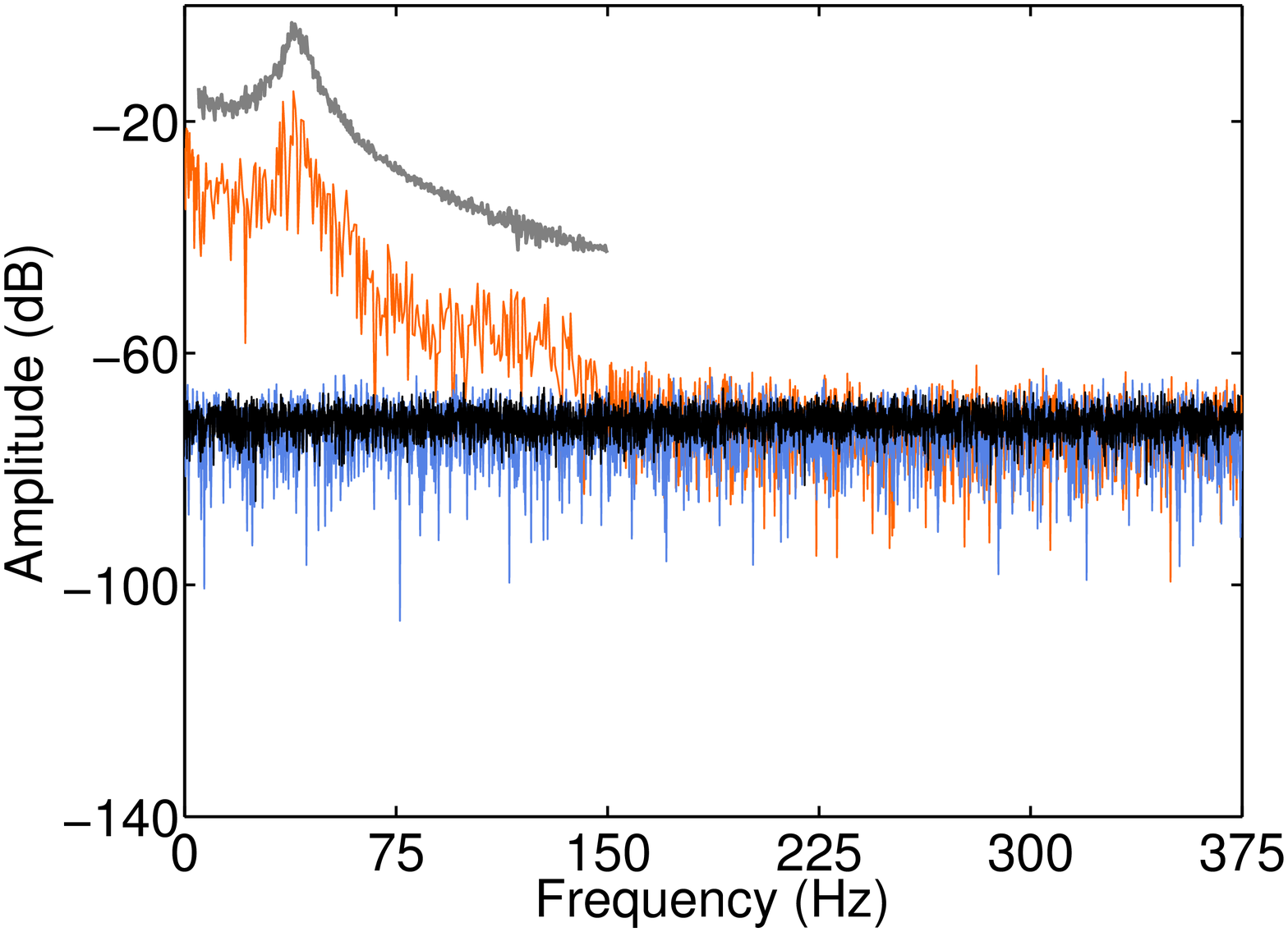}} \\
\end{tabular}
\caption{Nonparametric analysis of the nonlinear distortions affecting the Bouc-Wen system of Section~\ref{Sec:BoucWen}. The RMS input amplitude is equal to (a) 1, (b) 10, (c) 25 and (d) 50 $N$. Output level at the measurement lines (in grey); odd distortions (in orange); even distortions (in blue); noise level (in black).}
\label{Fig:NonParam_Results}
\end{center}
\end{figure}

The nonparametric analysis procedure is applied in Fig.~\ref{Fig:NonParam_Results} to the Bouc-Wen system of Section~\ref{Sec:BoucWen}. The excitation signal is a multisine with odd excited frequencies in 5 -- 150 $Hz$, and a frequency resolution $f_{0} = f_{s}/N \cong 0.09 \: Hz$, given $N = 8192$. Odd detection lines are created by randomly excluding one frequency in each group of 3 successive measurement lines. The RMS input amplitude is equal to 1, 10, 25 and 50 $N$ in Fig.~\ref{Fig:NonParam_Results}~(a -- d), respectively. The noise level displayed in black is obtained by averaging the measurements over 4 periods in steady state. Fig.~\ref{Fig:NonParam_Results} proves that the system features no even nonlinearity. Conversely, substantial odd distortions are detected, including at low forcing level in Fig.~\ref{Fig:NonParam_Results}~(a), where they lie 20 $dB$ below the output level in the resonance vicinity. At higher forcing levels in Fig.~\ref{Fig:NonParam_Results}~(b -- d), odd distortions affect the system response throughout the input band, in particular in the resonance frequency and the third harmonic regions. 

In summary, this section learns that identifying the Bouc-Wen system of Section~\ref{Sec:BoucWen} solely requires odd nonlinear model terms, which was retrieved by making exclusive use of output data. This is coherent with Eq.~(\ref{Eq:EOMBW_Hysteretic}) which writes, given the choice $\nu = 1$ in Table~\ref{Table:BWParams},
\begin{equation}
\dot{z}(y,\dot{y}) = \alpha\:\dot{y} - \beta \: \gamma \: \left|\dot{y}\right| z - \beta \: \delta \: \dot{y} \left|z\right| ,
\label{Eq:EOMBW_Hysteretic_nu1}
\end{equation}
where the expressions $\left|\dot{y}\right| z$ and $\dot{y} \left|z\right|$ are quadratic odd nonlinearities.

\newpage
\section{Nonlinear state-space identification}\label{Sec:PNLSS}

A nonlinear state-space model can be generally expressed in discrete-time form as
\begin{equation}
\left\lbrace
\begin{array}{r c l}
    \mathbf{x}(t+1) & = & \mathbf{g}\left( \mathbf{x}(t),\mathbf{u}(t),\boldsymbol{\theta} \right) \\
    \mathbf{y}(t) & = & \mathbf{h}\left( \mathbf{x}(t),\mathbf{u}(t),\boldsymbol{\theta} \right) ,
\end{array} \right.
\label{Eq:NLStateSpace}
\end{equation}
where $\mathbf{x} \in \mathbb{R}^{\: n}$ is the state vector, $\mathbf{u} \in \mathbb{R}^{\: q}$ the input vector, $\mathbf{y} \in \mathbb{R}^{\: l}$ the output vector, $\mathbf{g} \in \mathbb{R}^{\: n}$ and $\mathbf{h} \in \mathbb{R}^{\: l}$ two nonlinear functionals, and $n$ the model order. The vector $\boldsymbol{\theta} \in \mathbb{R}^{\: n_{\theta}}$ contains the parameters of the model to be estimated. The first relation in Eqs.~(\ref{Eq:NLStateSpace}) is known as the state equation, and dictates the dynamic evolution of the system. The second relation is the output equation, which translates the current system state and input into measurable output information.

\subsection{The polynomial nonlinear state-space model structure}\label{Sec:PNLSS_Model}

The nonlinear functionals $\mathbf{g}\left(\mathbf{x}(t),\mathbf{u}(t),\boldsymbol{\theta}\right)$ and $\mathbf{h}\left(\mathbf{x}(t),\mathbf{u}(t),\boldsymbol{\theta}\right)$ in Eqs.~(\ref{Eq:NLStateSpace}) can, in principle, be expanded using any basis functions. In this paper, a polynomial representation is adopted, following the original idea of Ref.~\cite{Paduart_PNLSS}. Polynomial expansions are attractive because they are simple, linear in their parameters, can be easily extended to the multivariate case, and possess universal approximation properties~\cite{PNLSS_Approximation}. Eqs.~(\ref{Eq:NLStateSpace}) become
\begin{equation}
\left\lbrace
\begin{array}{r c l}
    \mathbf{x}(t+1) & = & \mathbf{A} \: \mathbf{x}(t) + \mathbf{B} \: \mathbf{u}(t) + \mathbf{E} \: \mathbf{e}\left( \mathbf{x}(t),\mathbf{u}(t) \right) \\
    \mathbf{y}(t) & = & \mathbf{C} \: \mathbf{x}(t) + \mathbf{D} \: \mathbf{u}(t) + \mathbf{F} \: \mathbf{f}\left( \mathbf{x}(t),\mathbf{u}(t) \right) ,
\end{array} \right.
\label{Eq:PNLSS}
\end{equation}
where $\mathbf{A} \in \mathbb{R}^{\: n \times n}$, $\mathbf{B} \in \mathbb{R}^{\: n \times q}$, $\mathbf{C} \in \mathbb{R}^{\: l \times n}$ and $\mathbf{D} \in \mathbb{R}^{\: l \times q}$ are the linear state, input, output and feedthrough matrices, respectively. The vectors $\mathbf{e} \in \mathbb{R}^{\: n_{e}}$ and $\mathbf{f} \in \mathbb{R}^{\: n_{f}}$ include all monomial combinations of the state and input variables up to degree $d$. The associated coefficients are arranged in the matrices $\mathbf{E} \in \mathbb{R}^{\: n \times n_{e}}$ and $\mathbf{F} \in \mathbb{R}^{\: l \times n_{f}}$.

The number of nonlinear terms in Eqs.~(\ref{Eq:PNLSS}) is~\cite{Paduart_PNLSS}
\begin{equation}
s = \left( \frac{\left( n+q+d \right)!}{d! \: \left( n+q \right)!} -1 \right) \times \left( n+l \right) .
\label{Eq:NbrMonomials}
\end{equation}
This number can be reduced by probing the significance of each term in the decrease of the model error fit evaluated on validation data. In this respect, Ref.~\cite{JPaduart_Thesis} introduced several parsimonious alternatives to Eqs.~(\ref{Eq:PNLSS}). This includes considering nonlinear terms in the state equation only, disregarding input variables in the monomial combinations, or selecting non-consecutive polynomial degrees. These modelling strategies will be exploited in Section~\ref{Sec:PNLSS_Results} to avoid overfitting issues.

\subsection{Identification methodology}\label{Sec:PNLSS_ID}

A two-step methodology was proposed in Ref.~\cite{Paduart_PNLSS} to identify the parameters of the model structure in Eqs.~(\ref{Eq:PNLSS}). First, initial estimates of the linear system matrices $\left(\mathbf{A},\mathbf{B},\mathbf{C},\mathbf{D} \right)$ are calculated by measuring and fitting the best linear approximation (BLA) of the system under test. Second, assuming zero initial values for the nonlinear coefficients in $\left(\mathbf{E},\mathbf{F}\right)$, a full nonlinear model is built using optimisation.

\subsubsection{Initial linear model}\label{Sec:PNLSS_ID_Step1}

The BLA of a nonlinear system is defined as the linear model $G_{BLA}(j\omega_{k})$ which approximates best the system output in least-squares sense~\cite{JSchoukens_Book}. In general, it varies with the input frequency content and RMS value. The BLA can be measured by conducting $M$ experiments, consisting each in applying a multisine excitation and collecting $P$ steady-state periods of input-output data~\cite{JSchoukens_BookExercises}. In the single-input, single-output case, the frequency response function (FRF) associated with the $m$-th experiment is obtained as the ratio 
\begin{equation}
G_{m}(j\omega_{k}) = \frac{\displaystyle \frac{1}{P} \: \sum_{p=1}^{P} Y_{m,p}(k)}{\displaystyle \frac{1}{P} \: \sum_{p=1}^{P} U_{m,p}(k)} ,
\label{Eq:Experiment_m}
\end{equation}
where $U_{m,p}(k)$ and $Y_{m,p}(k)$ are the discrete Fourier transforms (DFTs) of the input $u(t)$ and output $y(t)$ acquired during the $p$-th period of the $m$-th experiment, respectively. The BLA is calculated as an averaged FRF over experiments, so that
\begin{equation}
G_{BLA}(j\omega_{k}) = \frac{1}{M} \displaystyle \sum_{m=1}^{M} G_{m}(j\omega_{k}) .
\label{Eq:BLA}
\end{equation}

A linear state-space model $\left(\mathbf{A},\mathbf{B},\mathbf{C},\mathbf{D} \right)$ is fitted to the nonparametric measurement of $G_{BLA}(j\omega_{k})$ in Eq.~(\ref{Eq:BLA}) using a frequency-domain subspace identification method~\cite{McKelvey_Subspace,Pintelon_Subspace}. The quality of the subspace model is evaluated through the weighted least-squares cost function 
\begin{equation}
V_{L} = \displaystyle \sum_{k=1}^{F} \epsilon_{L}^{H}(k) W_{L}(k) \: \epsilon_{L}(k) ,
\label{Eq:VL}
\end{equation}
where $F$ is the number of processed frequencies, $H$ denotes the Hermitian transpose, and $W_{L}(k)$ is a weighting function. Note that the proper selection of $W_{L}$ is studied in Section~\ref{Sec:PNLSS_Results}. The model fitting error $\epsilon_{L}(k)$ is defined as the difference
\begin{equation}
\epsilon_{L}(k) = G_{L}(j\omega_{k}) - G_{BLA}(j\omega_{k}) .
\label{Eq:EL}
\end{equation}
The transfer function of the linear subspace model is constructed as
\begin{equation}
G_{L}(j\omega_{k}) = \mathbf{C} \: \left(z_{k}\:\mathbf{I}^{n} - \mathbf{A}\right)^{-1} \: \mathbf{B} + \mathbf{D} ,
\label{Eq:GStateSpace}
\end{equation}
where $z_{k} = e^{j\left( 2\pi \:k/N \right)}$ is the z-transform variable and $\mathbf{I}^{n} \in \mathbb{R}^{\: n \times n}$ an identify matrix.

The subspace method of Ref.~\cite{McKelvey_Subspace} generally yields a reasonably low value of the cost function $V_{L}$. Minimising $V_{L}$ with respect to all parameters in $\left(\mathbf{A},\mathbf{B},\mathbf{C},\mathbf{D} \right)$ further improves the quality of the obtained linear model. As shown in Section~\ref{Sec:PNLSS_Results}, this also reduces its dependence upon an algorithmic dimensioning parameter $i$, which sizes the data matrices processed in the subspace identification~\cite{Paduart_PNLSS}. Moreover, the model order $n$ is, in practice, determined by carrying out the cost function minimisation for multiple $n$ values, and retaining the model with the lowest validation fitting error.

\subsubsection{Full nonlinear model}\label{Sec:PNLSS_ID_Step2}

The second step of the identification methodology involves minimising a second weighted least-squares cost function writing
\begin{equation}
V_{NL} = \displaystyle \sum_{k=1}^{F} \epsilon_{NL}^{H}(k) W_{NL}(k) \: \epsilon_{NL}(k) ,
\label{Eq:VNL}
\end{equation}
where $W_{NL}(k)$ is a weighting function discussed in Section~\ref{Sec:PNLSS_Results}. In Eq.~(\ref{Eq:VNL}), the error measure $\epsilon_{NL}(k)$ is defined as
\begin{equation}
\epsilon_{NL}(k) = Y_{NL}(k) - Y(k) ,
\label{Eq:ENL}
\end{equation}
where $Y_{NL}(k)$ and $Y(k)$ are the modelled and measured output DFT spectra, respectively. All parameters of the full nonlinear model $\left(\mathbf{A},\mathbf{B},\mathbf{C},\mathbf{D},\mathbf{E},\mathbf{F} \right)$ are estimated by minimising $V_{NL}$, starting from the linear system matrices obtained in Section~\ref{Sec:PNLSS_ID_Step1} and zero initial values for the nonlinear coefficients.

Note that the minimisation of the two cost functions in Eqs.~(\ref{Eq:VL}) and~(\ref{Eq:VNL}) is performed in this work using a Levenberg-Marquardt optimisation routine, which combines the large convergence region of the gradient descent method with the fast convergence of the Gauss-Newton method~\cite{Levenberg,Marquardt}. In this regard, technicalities related to the calculation of the Jacobian of Eqs.~(\ref{Eq:VL}) and~(\ref{Eq:VNL}) are elaborated in Ref.~\cite{Paduart_PNLSS}.

\newpage
\subsection{Identification results}\label{Sec:PNLSS_Results}

This section identifies the Bouc-Wen system of Section~\ref{Sec:BoucWen} using a polynomial nonlinear state-space model. A multisine excitation with all odd and even frequencies excited in the 5 -- 150 $Hz$ band is applied to synthesise 4 steady-state periods of measurement. The input amplitude level is fixed to 50 $N$ RMS, which leads to severe nonlinear effects, as visible in Fig.~\ref{Fig:NonParam_Results}~(d). We stress the deliberate choice to select two different input signals to perform the nonparametric analysis of output distortions and the parametric modelling of input-output data. In particular, a force spectrum including measurement and detection lines was required to distinguish odd from even nonlinearities in Section~\ref{Sec:NonParam} (see Fig.~\ref{Fig:NonParam_Results}), whereas a fully excited spectrum is utilised herein to capture the system dynamics over the complete band of interest.

To calculate the BLA of the Bouc-Wen system, 4 data sets are generated, \textit{i.e.} $M=4$, considering different realisations of the multisine phases and noise disturbances. The nonparametric estimate $G_{BLA}(j\omega_{k})$ calculated through Eq.~(\ref{Eq:BLA}) is transformed into a linear parametric state-space model $\left(\mathbf{A},\mathbf{B},\mathbf{C},\mathbf{D} \right)$ by applying subspace identification and subsequently minimising the cost function in Eq.~(\ref{Eq:VL}). The weighting function $W_{L}(k)$ is chosen as the inverse of the total variance of $G_{BLA}(j\omega_{k})$, hence encompassing the variability caused by nonlinear and noise distortions. This choice comes down to setting~\cite{Schoukens_VarianceBLA}
\begin{equation}
W_{L}(k) =  \frac{1}{M\left(M-1\right)} \displaystyle \sum_{m=1}^{M} \left| G_{m}(j\omega_{k}) - G_{BLA}(j\omega_{k}) \right|^{2} .
\label{Eq:WL}
\end{equation}

\begin{figure}[ht]
\begin{center}
\includegraphics[width=140mm]{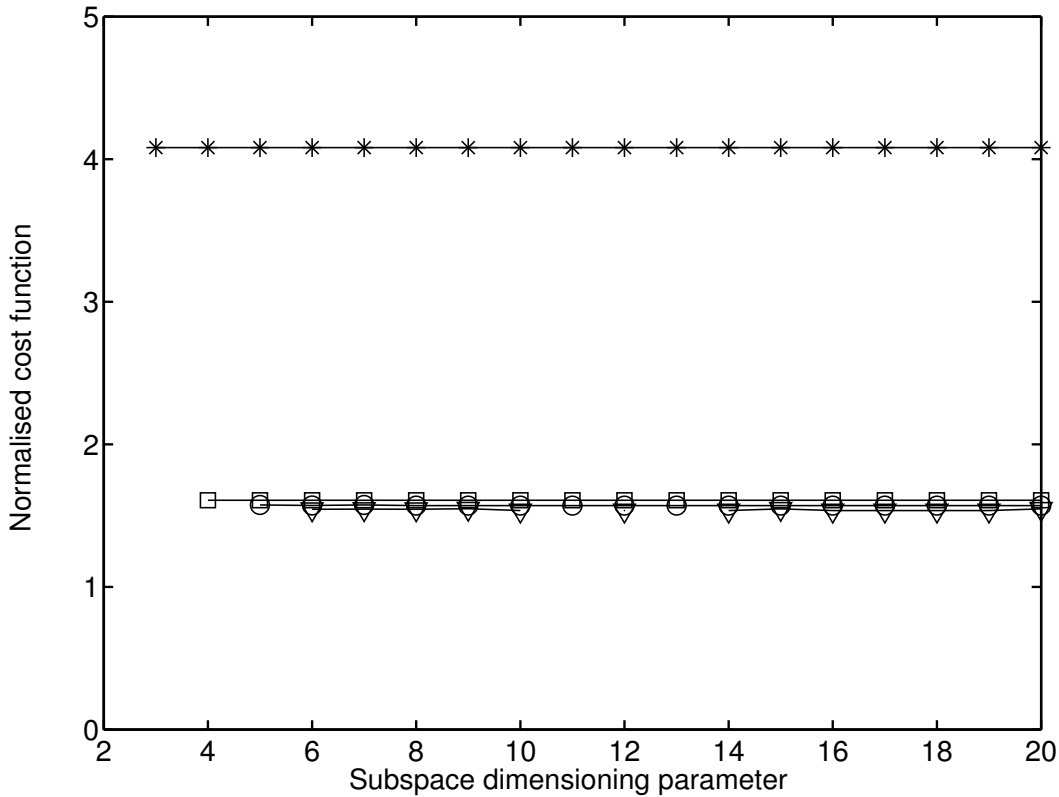}\\
\caption{Cost function $V_{L}$ normalised by the number of processed frequencies $F$ and plotted versus the subspace dimensioning parameter $i$. Model order 2 (stars), 3 (rectangles), 4 (circles) and 5 (triangles).}
\label{Fig:i_Modelorder}
\end{center}
\end{figure}

Fig.~\ref{Fig:i_Modelorder} depicts the minimised cost function $V_{L}$ versus the subspace dimensioning parameter $i$. The cost function is normalised by the number of processed frequencies $F = 1585$, and is plotted for model orders 2, 3, 4 and 5 using star, rectangular, circular and triangular labels, respectively. It is observed that $V_{L}$ is virtually insensitive to $i$, and that the order 3 corresponds to the best trade-off between model accuracy and parameter parsimony. Note that, for $n=3$, the minimum value of the cost function is obtained for $i=4$.

\begin{figure}[p]
\begin{center}
\begin{tabular}{c}
\subfloat[]{\label{Fig:Modelorder_2}\includegraphics[width=140mm]{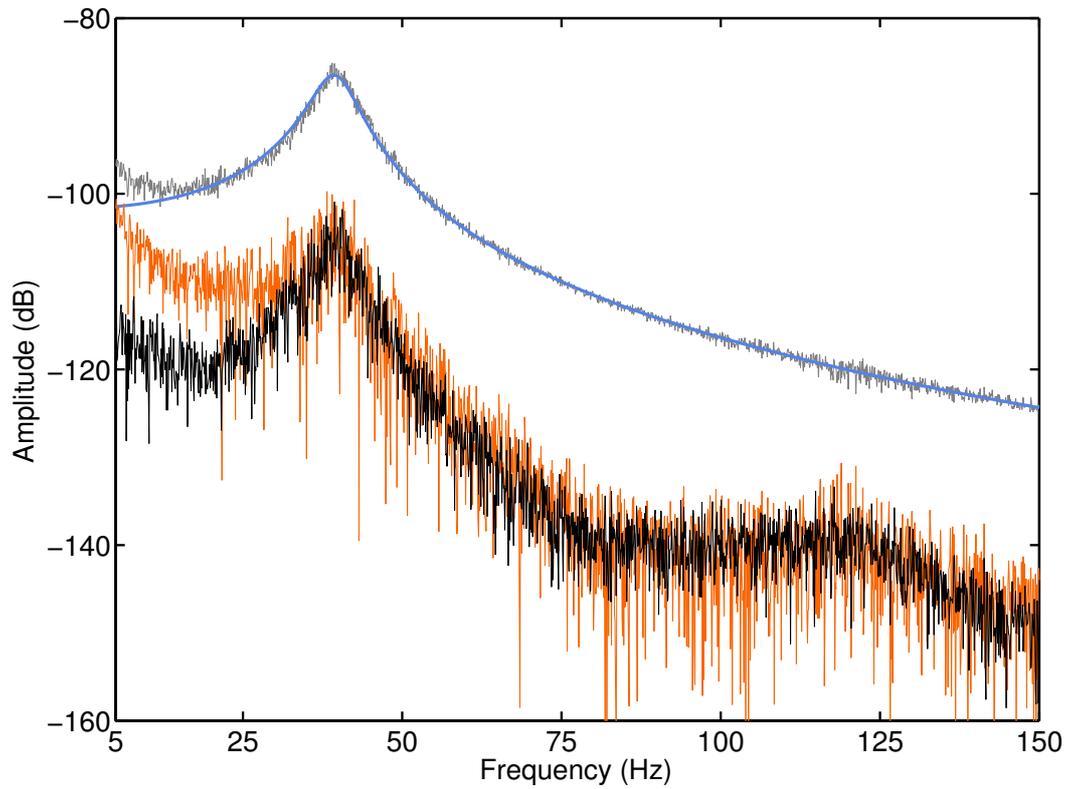}}\\
\subfloat[]{\label{Fig:Modelorder_3}\includegraphics[width=140mm]{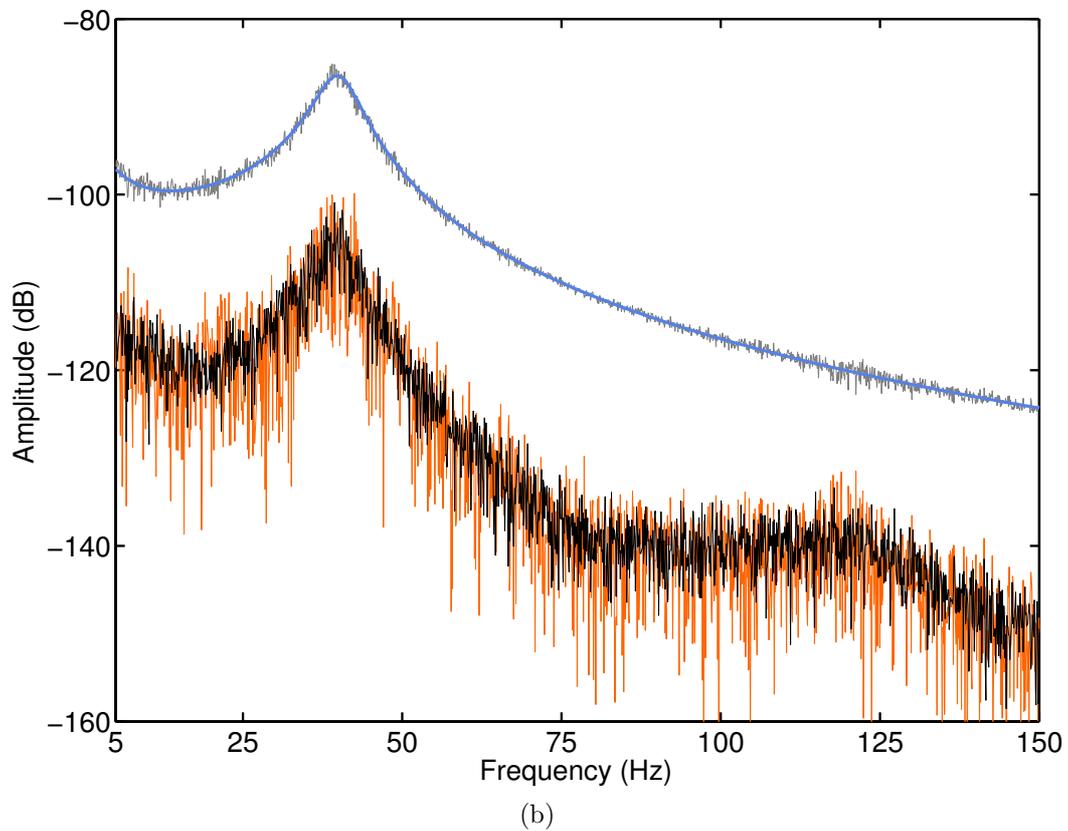}}\\
\end{tabular}
\caption{Nonparametric (in grey) and parametric (in blue) BLA, modelling error level (in orange) and total distortions level (in black). (a) $n=2$; (b) $n=3$.}
\label{Fig:Modelorder}
\end{center}
\end{figure}

The nonparametric and parametric BLA of the system are presented in Fig.~\ref{Fig:Modelorder} in grey and blue, respectively. An accurate fit based on a model of order 2 is achieved in Fig.~\ref{Fig:Modelorder}~(a), except for frequencies lower than 15 $Hz$. In this region, the modelling error level displayed in orange becomes substantially larger than the total distortions level plotted in black. By contrast, selecting $n=3$, as in Fig.~\ref{Fig:Modelorder}~(b), results in a perfect fitting throughout the input band, confirming the analysis of Fig.~\ref{Fig:i_Modelorder}. The need for a model of order 3 is substantiated by recasting the Bouc-Wen dynamics of Eqs.~(\ref{Eq:EOMBW}) and~(\ref{Eq:EOMBW_Hysteretic_nu1}) in the form
\begin{equation}
\left\lbrace
\begin{array}{r c l}
    \left(\begin{array}{c} \dot{y} \\
													 \ddot{y} \\
													 \dot{z} \\
					\end{array}	\right) & = &
					\left(\begin{array}{c c c} 0 & 1 & 0 \\
																		 -\frac{k_{L}}{m_{L}} & -\frac{c_{L}}{m_{L}} & -\frac{1}{m_{L}}	\\
																		 0 & \alpha & 0 \\
					\end{array}	\right)
					\: \left(\begin{array}{c} y \\
																	  \dot{y} \\
																		z \\
					\end{array}	\right)
					+ \left(\begin{array}{c} 0 \\
																	 \frac{1}{m_{L}}	\\
																	 0 \\
					\end{array}	\right)
					\: u
					+ \left(\begin{array}{c c} 0 & 0 \\
																	   0 & 0 \\
																	   -\beta \: \gamma & -\beta \: \delta \\
					\end{array}	\right)
					\: \left(\begin{array}{c} \left|\dot{y}\right| z \\
																	  \dot{y} \left|z\right| \\																		
					\end{array}	\right) \\
		& & \\
    y & = & \left(\begin{array}{c c c} 1 & 0 & 0 \\
					\end{array}	\right)
					\: \left(\begin{array}{c} y \\
																	  \dot{y} \\
																		z \\
					\end{array}	\right) .
\end{array} \right.
\label{Eq:SDOFBW_StateSpace}
\end{equation}
Eqs.~(\ref{Eq:SDOFBW_StateSpace}) show that translating the Bouc-Wen equations in state space requires the definition of 3 state variables. It should also be noted that the state matrix $\mathbf{A}$ identified for $n=3$ possesses a pair of complex conjugate poles and one real pole. The appearance in the fitted model of a real pole, \textit{i.e.} pole with zero frequency, is consistent with the definition of hysteresis as a quasi-static phenomenon, as explained in Section~\ref{Sec:Introduction}. 

\begin{figure}[p]
\begin{center}
\includegraphics[width=140mm]{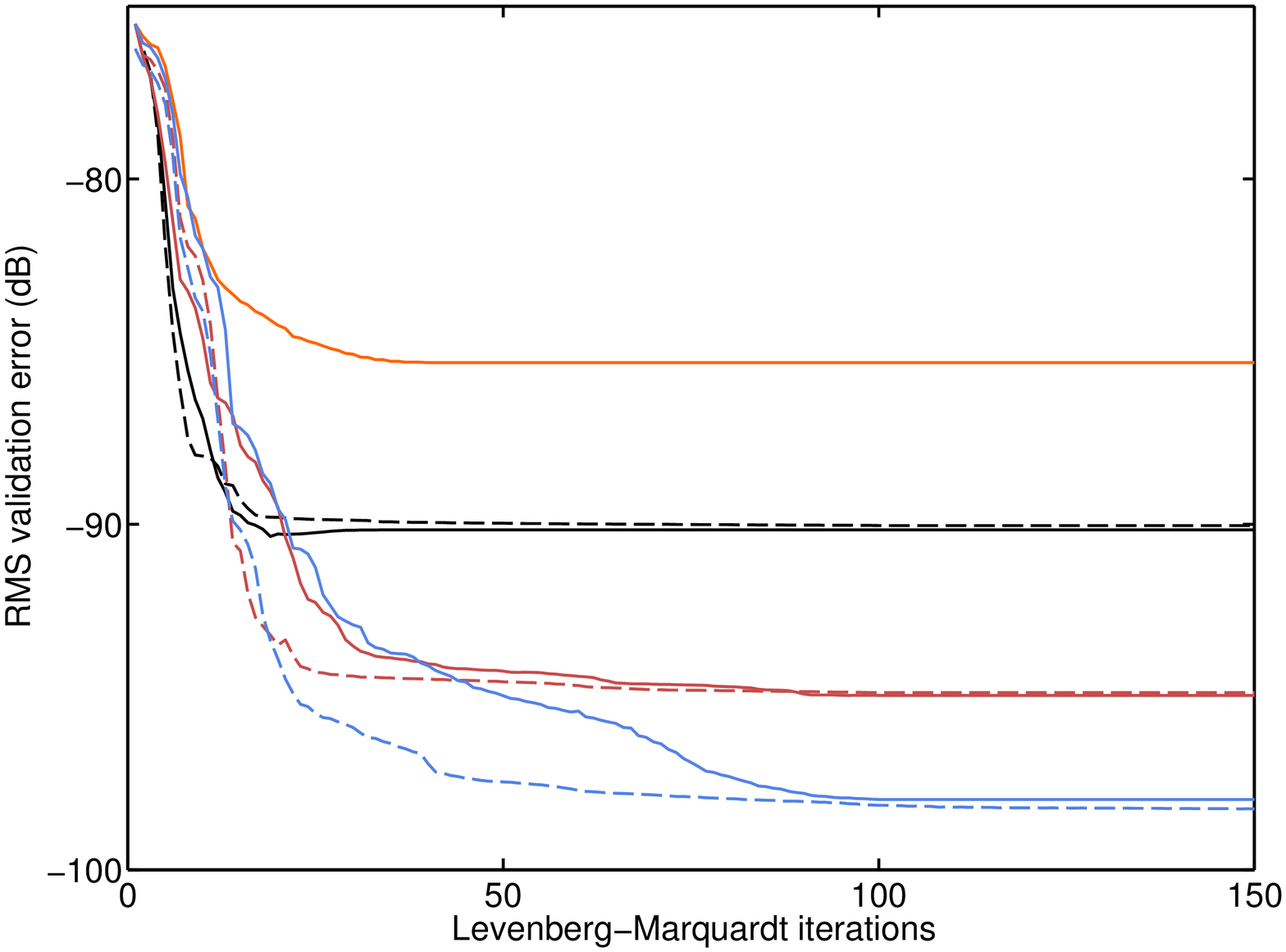}\\
\caption{Decrease of the RMS validation error over 150 Levenberg-Marquardt iterations. Polynomial nonlinear state-space models of degree 2 (in orange), 2--3 (in solid black), 2--3--4 (in dashed black), 2--3--4--5 (in solid red), 2--3--4--5--6 (in dashed red), 2--3--4--5--6--7 (in solid blue), and 3--5--7 (in dashed blue).}
\label{Fig:LMiterations}
\end{center}
\end{figure}

Final estimates of all nonlinear state-space parameters $\left(\mathbf{A},\mathbf{B},\mathbf{C},\mathbf{D},\mathbf{E},\mathbf{F} \right)$ are obtained by minimising the cost function $V_{NL}$. A unit weighting $W_{NL}(k)$ is applied in Eq.~(\ref{Eq:VNL}), reflecting that unmodelled dynamics, which is assumed to be uniformly distributed in the frequency domain, dominates noise disturbances. Fig.~\ref{Fig:LMiterations} plots the decrease of the RMS modelling error over 150 Levenberg-Marquardt iterations. This error is evaluated on a validation data set generated considering the same excitation properties as for estimation data. The converged value of the error for different nonlinear models is given in Table~\ref{Table:PNLSS}, together with their respective number of parameters. As a result of the odd nature of the nonlinearities in Eqs.~(\ref{Eq:EOMBW_Hysteretic_nu1}) and~(\ref{Eq:SDOFBW_StateSpace}), it is found that introducing in the model even-degree monomials brings no decrease of the validation error, confirming the physical intuition gained in Section~\ref{Sec:NonParam}. The most accurate state-space model reported in Table~\ref{Table:PNLSS} comprises odd monomials up to $d=7$, for a total of 217 parameters. Note that all the models in Fig.~\ref{Fig:LMiterations} and Table~\ref{Table:PNLSS} do neither incorporate nonlinear terms in the output equation, nor input variables in the monomial combinations, in accordance with Eqs.~(\ref{Eq:SDOFBW_StateSpace}). 

\begin{table}[p]
\centering
\begin{tabular*}{1.00\textwidth}{@{\extracolsep{\fill}} l c c}
\hline
Polynomial degree & RMS validation error ($dB$) & Number of parameters \\
 & & \\
2 & -85.32 & 34 \\
2-3 & -90.35 & 64 \\
2-3-4 & -90.03 & 109 \\
2-3-4-5 & -94.87 & 172 \\
2-3-4-5-6 & -94.85 & 256 \\
2-3-4-5-6-7 & -97.96 & 364 \\
3-5-7 & -98.32 & 217 \\
\hline
\end{tabular*}
\caption{RMS validation error for polynomial nonlinear state-space models of various degrees together with their respective number of parameters.} 
\label{Table:PNLSS}
\end{table}

\begin{figure}[p]
\vspace*{-2cm}
\begin{center}
\begin{tabular}{c}
\subfloat[]{\label{Fig:OutputFFT_23}\includegraphics[width=140mm]{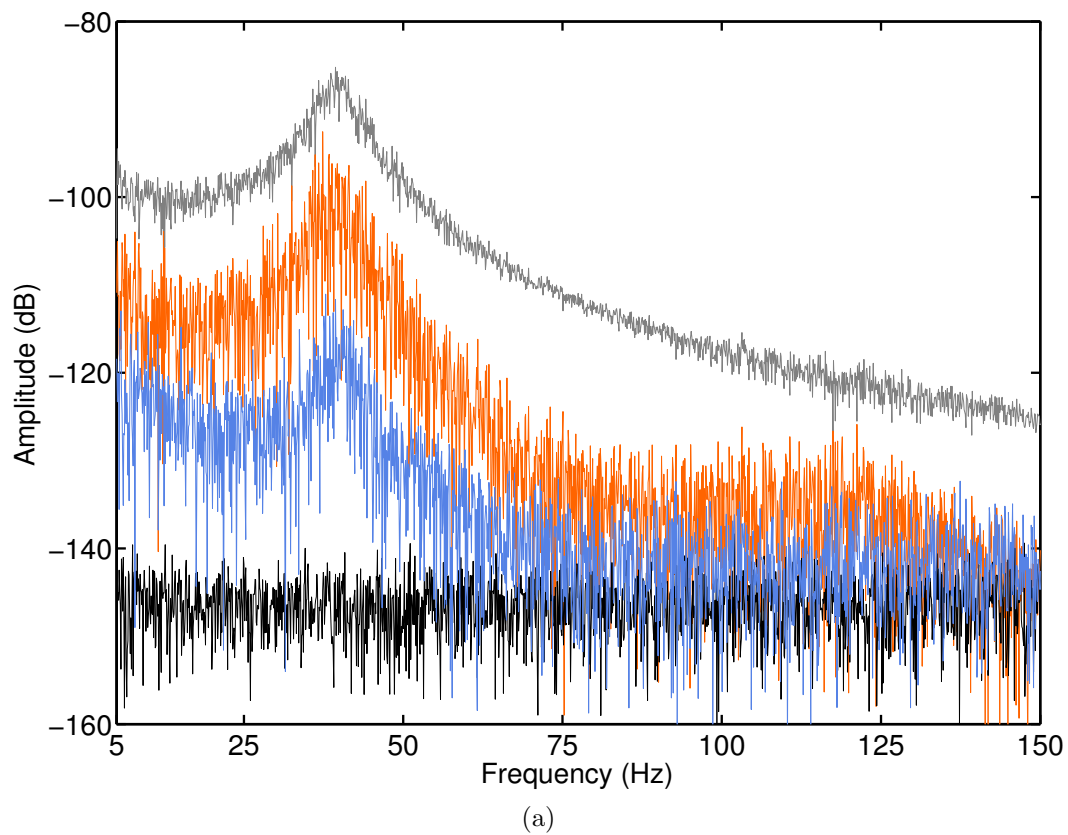}}\\
\subfloat[]{\label{Fig:OutputFFT_357}\includegraphics[width=140mm]{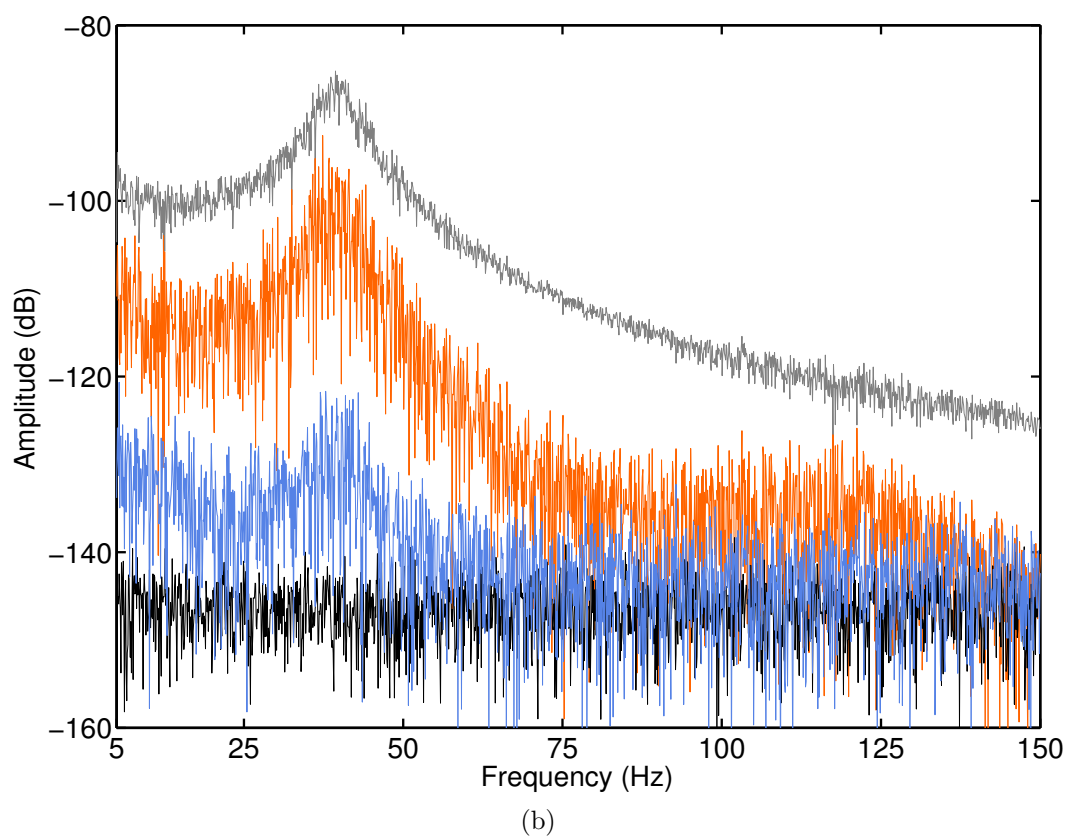}}\\
\end{tabular}
\caption{Frequency-domain behaviour of the validation modelling error over the input band, featuring the output spectrum (in grey), the linear (in orange) and nonlinear (in blue) fitting error levels, and the noise level (in black). (a) Monomials of degree 2 and 3; (b) monomials of degree 3, 5 and 7.}
\label{Fig:OutputFFT}
\end{center}
\end{figure}

\begin{figure}[ht]
\begin{center}
\includegraphics[width=140mm]{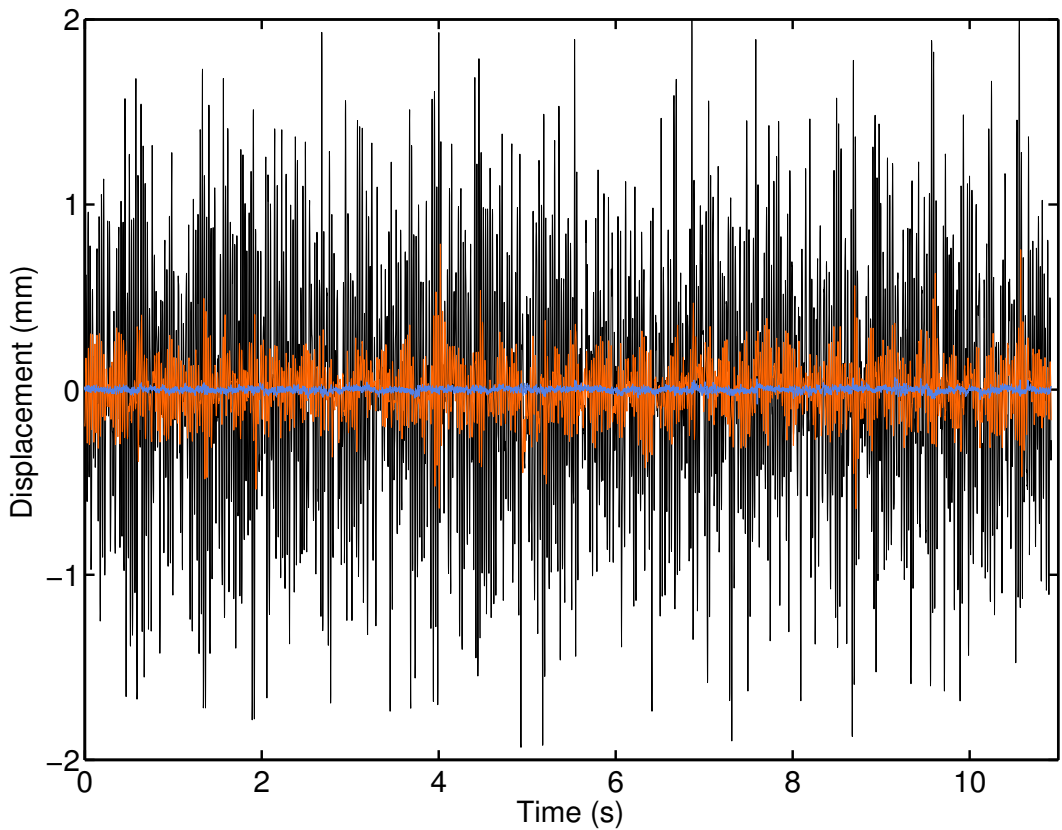}\\
\caption{Time-domain behaviour of the validation modelling error for monomials of degree 3, 5 and 7, featuring the output time history (in black) and the linear (in orange) and nonlinear (in blue) fitting error levels.}
\label{Fig:Output_2357}
\end{center}
\end{figure}

\newpage
The frequency-domain behaviour of the validation modelling error is studied in Fig.~\ref{Fig:OutputFFT}, where the output spectrum in grey is compared with linear and nonlinear fitting error levels in orange and blue, respectively. Using monomials of degree 2 and 3 in the state variables, as is the standard choice~\cite{Paduart_PNLSS}, reduces the linear error by a factor of 20 $dB$, as visible in Fig.~\ref{Fig:OutputFFT}~(a). A further decrease of 10 $dB$ is achieved by selecting monomials of degree 3, 5 and 7, as in Fig.~\ref{Fig:OutputFFT}~(b). It should be remarked that an exact polynomial description of the nonlinearities $\left|\dot{y}\right| z$ and $\dot{y} \left|z\right|$ in Eqs.~(\ref{Eq:EOMBW_Hysteretic_nu1}) demands, in principle, an infinite series of terms, preventing the nonlinear errors in Fig.~\ref{Fig:OutputFFT} from reaching the noise level depicted in black. Increasing the polynomial degree $d$ to values higher than 7, though being manifestly possible, was not attempted in this work to limit the computational burden involved in the cost function minimisation. Finally, the time-domain errors corresponding to Fig.~\ref{Fig:OutputFFT}~(b) are depicted in Fig.~\ref{Fig:Output_2357}. The RMS values of the validation output time history and of the linear and nonlinear errors are equal to 0.66, 0.15 and 0.01 $mm$, respectively. This graph nicely illustrates the important increase of the identification accuracy obtained by introducing nonlinear black-box terms in the state-space modelling of hysteresis.

\newpage
\section{Model validation under sine-sweep excitations}\label{Sec:Validation}

\begin{figure}[ht]
\begin{center}
\includegraphics[width=140mm]{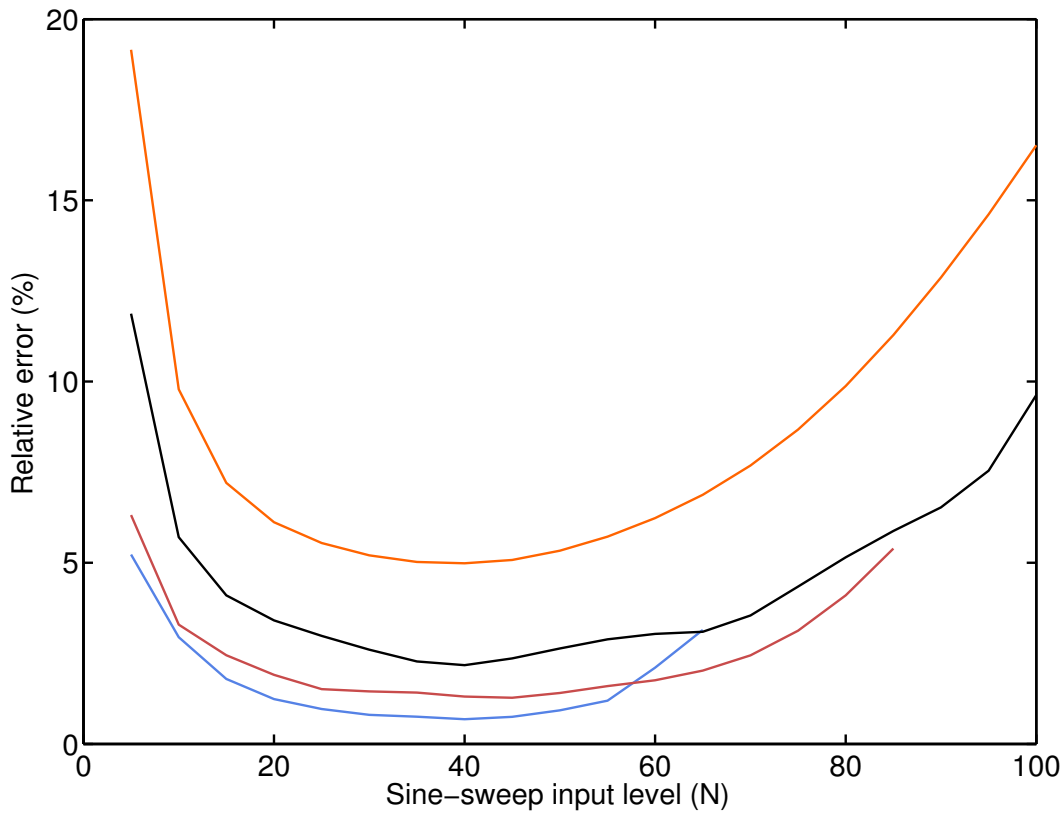} \\
\caption{Relative error (in $\%$) between the exact and reconstructed responses of the Bouc-Wen system for sine-sweep forcing amplitudes ranging from 5 to 100 $N$. Nonlinear state-space models with monomials of degree 2 (in orange), 2--3 (in black), 2--3--5 (in red), 2--3--5--7 (in blue).}
\label{Fig:SW_RelE_Models}
\end{center}
\end{figure}

This final section investigates the domain of validity of the state-space models fitted in Section~\ref{Sec:PNLSS_Results} under sine excitation signals. In particular, a comparison is made between the exact and reconstructed responses of the Bouc-Wen system under various sine-sweep forcing levels. Fig.~\ref{Fig:SW_RelE_Models} presents the relative error in percent between the two responses for input amplitudes ranging from 5 to 100 $N$. Four different nonlinear state-space models are analysed in this figure, namely comprising monomials of degree 2 (in orange), 2--3 (in black), 2--3--5 (in red) and 2--3--5--7 (in blue). The chosen input signals sweep the interval from 20 to 50 $Hz$ at a linear rate of 10 $Hz/min$. Reconstructed outputs are simulated in discrete time, with zero initial conditions, by evaluating Eqs.~(\ref{Eq:PNLSS}), which are explicit relations in $\mathbf{y}(t)$. It is observed that, as the model complexity increases, the prediction capabilities improve. The minimum relative error is achieved for all tested models around 40 $N$, reaching 0.7 $\%$ in the case of the 2--3--5--7 model.

However, complex models are likely to become unstable when extrapolated outside their fitting domain. This is visible for the blue and red models in Fig.~\ref{Fig:SW_RelE_Models}, which do no longer predict bounded outputs for input levels higher than 65 and 85 $N$, respectively. It is reminded that all models were estimated using a multisine excitation with a RMS value equal to 50 $N$ and all odd and even frequencies excited in the 5 -- 150 $Hz$ band.

\begin{figure}[ht]
\vspace{1cm}
\begin{center}
\includegraphics[width=140mm]{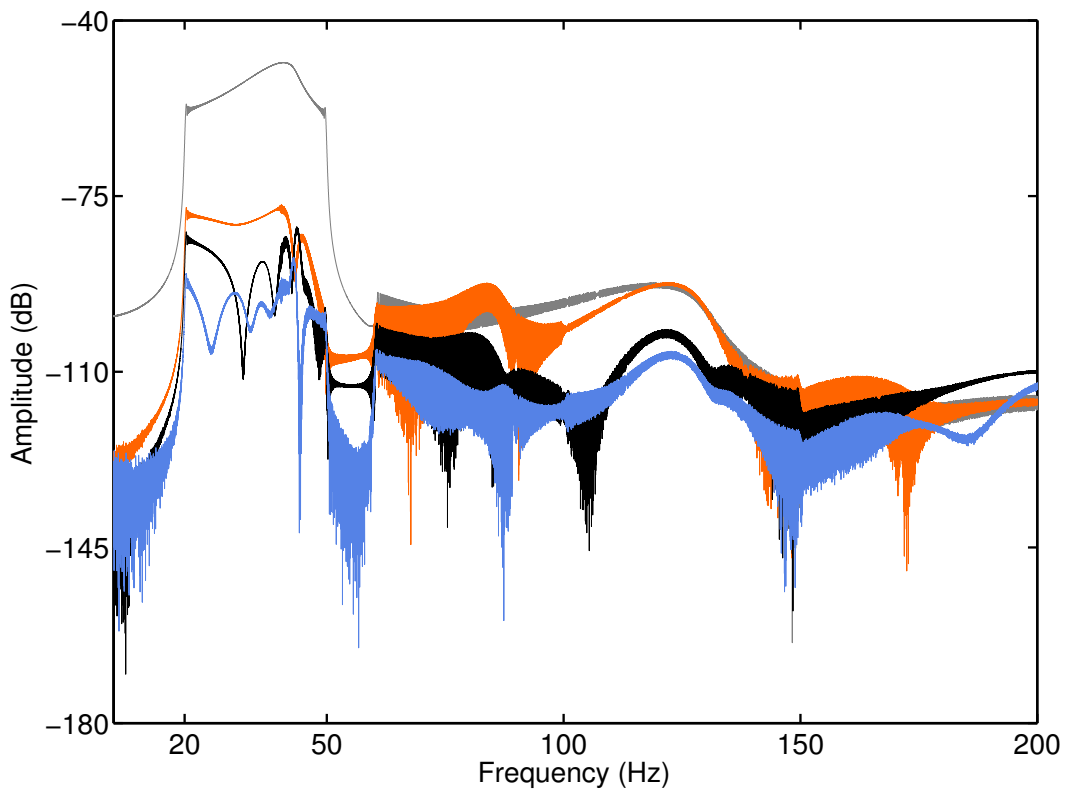} \\
\caption{Exact output spectrum over 5 -- 200 $Hz$ at 40 $N$ sine-sweep level (in grey) and reconstruction error for the 2 (in orange), 2--3 (in black) and 2--3--5--7 (in blue) nonlinear state-space models.}
\label{Fig:SW_OutputFFT}
\end{center}
\end{figure}

A frequency-domain error analysis is performed at 40 $N$ input level in Fig.~\ref{Fig:SW_OutputFFT}. The exact output spectrum is plotted in grey, and is compared with the reconstruction error for the 2 (in orange), 2--3 (in black) and 2--3--5--7 (in blue) state-space models. The error is seen to decrease in the input band for increasing model complexity. In the vicinity of the third harmonic around 120 $Hz$, the error similarly drops for higher polynomial degrees. Note that the cut-off frequency of 150 $Hz$ of the multisine excitation under which the state-space models were fitted is visible in the error plots of Fig.~\ref{Fig:SW_OutputFFT}.

\FloatBarrier

\newpage
\section{Conclusions}\label{Sec:Conclusion}

The purpose of the present paper was to introduce a general framework to identify hysteresis in dynamic systems. State-space models with polynomial nonlinear terms were selected to support this framework. They are fitted to data using a rigorous two-step methodology involving weighted least-squares minimisation. A numerical study was conducted to demonstrate the fitting accuracy of the proposed approach. The identified black-box models were also found to be reasonably parsimonious, given that they require no a priori knowledge about the observed hysteretic behaviour. This paper paves the way for addressing the experimental modelling of hysteresis featured in real applications, especially in the dynamics of jointed structures.

However, in the case of structures with multiple modes and multiple hysteretic components, the great number of parameters involved in the construction of the multivariate polynomials in Eq.~(\ref{Eq:PNLSS}) may become a limitation. The recent contribution by Dreesen and co-authors~\cite{Dreesen_DecouplingPNLSS} proposes to prune the nonlinear model parameters in state-space identification by employing advanced tensor techniques. It specifically demonstrates that the functions $\mathbf{e}\left( \mathbf{x}(t),\mathbf{u}(t) \right)$ and $\mathbf{f}\left( \mathbf{x}(t),\mathbf{u}(t) \right)$ in Eq.~(\ref{Eq:PNLSS}) can be replaced with decoupled polynomial representations. Formally, concatenating $\mathbf{x}$ and $\mathbf{u}$ as
\begin{equation}
\mathbf{\xi}(t) = \left( \mathbf{x}(t)^{T} \ \mathbf{u}(t)^{T} \right)^{T} ,
\label{Eq:xu}
\end{equation}
where $T$ denotes the matrix transpose operation, a simplified representation of $\mathbf{e}\left( \mathbf{x}(t),\mathbf{u}(t) \right)$ (a similar reasoning holds for $\mathbf{f}\left( \mathbf{x}(t),\mathbf{u}(t) \right)$) writes
\begin{equation}
\mathbf{e}\left( \mathbf{x}(t),\mathbf{u}(t) \right) = \mathbf{W} \mathbf{d}\left(\mathbf{V}^{T} \mathbf{\xi}(t) \right),
\label{Eq:DecE}
\end{equation}
where $\mathbf{V}$ and $\mathbf{W}$ are linear transformation matrices, and where $\mathbf{d}$ is a multivariate, decoupled polynomial function, \textit{i.e.} each component of $\mathbf{d}$ is a single-input, single-output function. The integration of this decoupling strategy into the hysteresis identification framework proposed herein is currently under investigation.

\section*{Acknowledgements}

This work was supported in part by the Fund for Scientific Research (FWO-Vlaanderen), by the Flemish Government (Methusalem), by the Belgian Government through the Inter university Poles of Attraction (IAP VII) Program, and by the ERC advanced grant SNLSID, under contract 320378.

\bibliographystyle{unsrt}
\bibliography{MSSP_BoucWenID_Bibliography}

\end{document}